\documentclass[floatfix,aps,preprint]{revtex4-1}
\usepackage{mathrsfs}
\usepackage{amsfonts}
\usepackage{amsmath}
\usepackage{amssymb}
\usepackage{revsymb}
\usepackage{graphicx}
\usepackage{bm}
\newcommand{\be}{\begin{equation}}
\newcommand{\ee}{\end{equation}}
\newcommand{\ba}{\begin{array}}
\newcommand{\ea}{\end{array}}
\newcommand{\bea}{\begin{eqnarray}} 
\newcommand{\eea}{\end{eqnarray}} 
\newcommand{\bd}{\begin{displaymath}}
\newcommand{\ed}{\end{displaymath}}
\newcommand{\bc}{\begin{center}}
\newcommand{\ec}{\end{center}}

\newcommand{\uveceps}{\hat{\pmb{\varepsilon}}}
\newcommand{\uvec}[1]{\hat{\mbf{#1}}}
\newcommand{\eps}{\varepsilon}
\newcommand{\mbf}[1]{\mathbf{#1}}

\newcommand{\trm}[1]{\textrm{#1}}

\newcommand{\vphi}{\varphi}
\newcommand{\sech}{\trm{sech}}
\newcommand{\ttau}{\tilde{\tau}}

\long\def\symbolfootnote[#1]#2{\begingroup%
\def\thefootnote{\fnsymbol{footnote}}\footnote[#1]{#2}\endgroup}

\newcommand{\figref}[1]{Fig. \ref{#1}}

\newcommand{\eqnref}[1]{Eq. (\ref{#1})}
\newcommand{\eqnrefs}[2]{Eqs. (\ref{#1}) and (\ref{#2})}
\newcommand{\eqnreft}[2]{Eqs. (\ref{#1}-\ref{#2})}

\newcommand{\tabref}[1]{Tab.$~$\ref{#1}}

\fussy

\fussy

\begin{document}
\title{Photon-photon scattering in collisions of laser pulses}
\author{B. \surname{King}}
  \email{ben.king@physik.uni-muenchen.de}
  \affiliation{Max-Planck-Institut f\"ur Kernphysik,
    Saupfercheckweg 1, D-69117 Heidelberg, Germany}
\affiliation{Ludwig-Maximilians-Universit\"at M\"unchen,
    Theresienstra\ss e 37, 80333 M\"unchen, Germany}
\author{C.~H. \surname{Keitel}}
  \email{keitel@mpi-hd.mpg.de}
  \affiliation{Max-Planck-Institut f\"ur Kernphysik,
    Saupfercheckweg 1, D-69117 Heidelberg, Germany}

\date{\today}
\begin{abstract}
A scenario for measuring the predicted processes of vacuum elastic and inelastic
photon-photon scattering with modern lasers is investigated. Numbers of
measurable scattered photons are calculated for the collision of two,
Gaussian-focused, pulsed lasers. We show that a single $10~\trm{PW}$ optical
laser beam split into two counter-propagating pulses is sufficient for measuring
the elastic process. Moreover, when these pulses are sub-cycle, our results
suggest the inelastic process should be measurable too.
\end{abstract}

\maketitle

\section{Introduction}
Quantum electrodynamics is commonly regarded to be a fantastically successful
theory whose accuracy has been tested to one part in $10^{12}$ for free
electrons \cite{gabrielse06} and one part in $10^{9}$ for bound electrons
\cite{sturm11}. However, among its several predictions that have yet to be
confirmed, is the nature of electromagnetic interaction with the quantised
vacuum. Already with the pioneering work of Sauter, Heisenberg and Euler
\cite{sauter31, heisenberg_euler36}, it was clear that quantum mechanics
predicts how particles traversing the classically empty space of the vacuum can
interfere with ephemeral ``virtual'' quantum states, whose lifetimes are of
durations permitted by the uncertainty relation. Virtual electron-positron
pairs, can in principle, be polarised by an external electromagnetic field, thus
introducing non-linearities into Maxwell's equations, which break the familiar
principle of superposition of electromagnetic waves in vacuum. Photons from
multiple, vacuum-polarising sources, can then become coupled on the common
point of interaction of the polarised virtual pairs. This process is predicted
to manifest itself in a variety of ways such as in a phase shift in intense
laser beams crossing one another \cite{tommasini07}, in a frequency shift of a
photon propagating in an intense laser \cite{mendonca06}, in polarisation
effects in crossing lasers such as vacuum birefringence and dichroism
\cite{king10b, dipiazza_PRL_06, heinzl_birefringence06}, where ideas have
already found an applied formulation \cite{tajima11a}, in dispersion effects
such as vacuum diffraction \cite{king10a, hatsagortsyan11} and also in vacuum
high harmonic generation \cite{dipiazza_harmonic05}. The typical scale for such
``refractive'' vacuum polarisation effects, where no pair-creation takes place,
is given by the critical field strength required to ionise a virtual
electron-positron pair, namely the pair-creation scale of $E_\trm{cr} =
m^{2}c^{3}/e\hbar = 1.3\times10^{16}~\trm{Vcm}^{-1}$ or an equivalent critical
intensity of $I_{\trm{cr}}=2.3\times10^{29}~\trm{Wcm}^{-2}$,
where $m$ and $-e<0$ are the mass and charge of an electron respectively.
Although this intensity lies some seven orders of magnitude above the record
high produced by a laser \cite{yanovsky08}, recent progress at facilities such
as the ongoing $10~\trm{PW}$ upgrade to the Vulcan laser \cite{vulcan_site} as
well as proposals for next generation lasers HiPER and ELI aiming at three
to four orders of magnitude less than critical, will put the experimental
verification of these long-predicted non-linear vacuum polarisation effects
finally within reach. This therefore motivates more realistic quantitative
predictions.
\newline


In the current paper, we focus on the phenomenon of photon-photon scattering,
which can either be elastic in the sense of a diffractive effect, or inelastic,
in the sense of four-wave mixing, allowing the frequency of one field to be
shifted up or down in multiples of the frequency of the others. When all
external fields have the same frequency, four-wave mixing is then equivalent to
lowest order vacuum high-harmonic generation. As an elastic process, numbers of
scattered photons have been calculated in the passage of one monochromatic
Gaussian laser beam through another \cite{tommasini10}, as well as in so-called
single- and ``double-slit'' set-ups \cite{king10a, king10b}, where a probe
Gaussian beam meets two other intense ones. Inelastic photon-photon scattering
has been investigated theoretically as a four-wave mixing process using
TE$_{10}$ and TE$_{01}$ modes in a superconducting cavity
\cite{marklund_PRL_01}, in the collision of three, perpendicular, plane-waves
\cite{lundstroem_PRL_06}
and as generating odd harmonics involving a single, spatially-focused
monochromatic wave \cite{fedotov_harmonics06}.  By incorporating both the pulsed
and spatially-focused nature of modern high-intensity laser beams, we perform a
more accurate calculation of the signal of the elastic scattering process. We
thereby investigate the robustness of the effect with a more detailed
calculation than hitherto performed, including dependency on beam collision
angle, impact parameter (lateral beam separation), longitudinal phase difference
(through lag) and pulse duration (finite beam length). Inclusion of four-wave mixing terms with a
pulsed set-up allows us, moreover, to determine the possibility of measuring
inelastic photon-photon scattering when a single $10~\trm{PW}$ beam is split
into two counter-propagating sub-cycle pulses.  In what follows, we work in
Gaussian cgs units (fine-structure constant $\alpha = e^{2}$), with $\hbar = c =
4\pi\varepsilon_{0} = 1$, unless explicit units denote otherwise.



\section{Scenario considered}
\begin{figure}[!ht]
\noindent\centering
\includegraphics[draft=false,
width=0.65\linewidth]{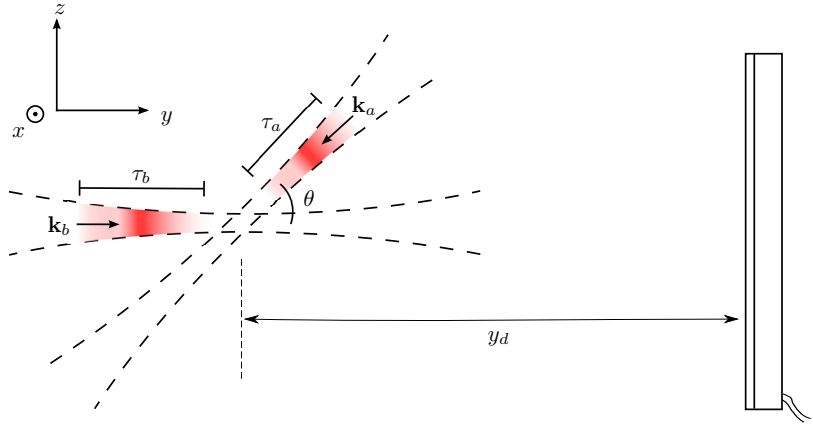}
\caption{The envisaged experimental set-up. $\tau_{a,b}$ refer to the pulse durations in
the Gaussian beam envelopes $\mbox{e}^{-(t\pm y)^{2}/\tau_{a,b}^{2}}$, the $\mbf{E}_{b}$ beam is displaced from the $y$-axis by co-ordinates
$x_0, z_0$, both beams have in general a carrier-envelope phase and the
$\mbf{E}_{a}$ beam lags behind $\mbf{E}_{b}$ by $\Delta t$. The $\mbf{E}_{b}$
field is incident on the detector.\label{fig:exp_setup}} 
\end{figure}

In order to analyse the collision of two laser pulses, several collision
parameters have been included. The envisaged scenario is shown in
\figref{fig:exp_setup}, in addition to which, lateral and temporal centring and
carrier envelope phase appear in the analytical set-up. Spatial focusing and
temporal pulse shape are present in taking the leading order spatial and
temporal terms of the Gaussian beam solution to Maxwell's equations (see e.g.
\cite{salamin_review06}). These approximations neglect terms of the order
$O(w_{c,0}/y_{r,c})$ and $O(1/\omega_{c} \tau_{c})$ respectively, where $c \in
\{a,b\}$ is used throughout for beams $a$ and $b$, the minimum beam waist is
$w_{c,0}$, Rayleigh length $y_{r,c} = \omega_{c}w_{c,0}^{2}/2$, beam frequency
$\omega_{c}$ and full-width-half-max pulse duration $\tau_{\trm{FWHM}}$ related
to $\tau$ via $\tau \sqrt{2\ln 2} = \tau_{\trm{FWHM}}$. The condition
$\omega_{c} \tau_{c} \gg 1$ limits the minimum pulse duration that can
be consistently considered in our analysis. For the electric fields of the two beams
$\mbf{E}_{a}$, $\mbf{E}_{b}$, we then have:
\bea
\mbf{E}_{a}(x,y,z,t) &=& \hat{\pmb{\eps}}_{a}'
\frac{E_{a,0}\,\mbox{e}^{-\frac{x'^{2}+z'^{2}}{w_{a}^{2}(y')}}}{\sqrt{1+(y'/y_{r
,a})^{2}}} \sin\left[\psi_{a} + \omega_{a}(t-\Delta t + y') -
\eta_{a}(y')\right]f_{a}(t-\Delta t+y') \label{eqn:Ea}\\
\mbf{E}_{b}(x,y,z,t) &=& \hat{\pmb{\eps}}_{b}
\frac{E_{b,0}\,\mbox{e}^{-\frac{x^{2}+z^{2}}{w_{b}^{2}(y)}}}{\sqrt{1+(y/y_{r,b}
)^{2}}} \sin\left[\psi_{b} + \omega_{b}(t-y) +\eta_{b}(y)
\right]f_{b}(t-y)\label{eqn:Eb}\\
\eta_{c}(y) &=& \tan^{-1}\left(\frac{y}{y_{r,c}}\right) -
\frac{\omega_{c}y}{2}\frac{x^{2}+z^{2}}{y^{2}+y^{2}_{r,c}}
\eea
where the co-ordinates $(x,y',z')$ are the same as $(x,y,z)$ rotated
anti-clockwise around the $x$ axis by an angle $\theta$,
with the polarisation $\hat{\pmb{\eps}}_{a}'$ being similarly rotated so that
$\mbf{k}_{c}\cdot\,\hat{\pmb{\eps}}_{c}=\mbf{k}_{c}'\cdot\,\hat{\pmb{\eps}}_{c}'=0$ and $|\hat{\pmb{\eps}}_{c}|=|\hat{\pmb{\eps}}_{c}'|=1$, where
$\mbf{k}_{c}$ is the beam wavevector, $f_{c}$ describes the pulse shape with
$f_{c}(x) = \mbox{e}^{-(x/\tau_{c})^{2}}$ being used, $w_{c}$ is the beam waist
$w^{2}_{c} = w_{c,0}^{2}\left(1+(y/y_{r,c})^{2}\right)$ dependent on transverse
co-ordinate, $\psi_{c}$ is a constant phase, $\Delta t$ is the lag and $E_{c,0}$ is the field amplitude, which satisfies $\int
dt\,dx\,dz\, |\mbf{E}(x,y=0,z,t)|^{2}/(4\pi) = \mathcal{E}$, with total beam
energy $\mathcal{E}$, or $E_{c,0} = 2\sqrt{2 P_{c,0}}/w_{c,0}$, for peak beam
power $P_{c,0}$, where we have already assumed that corrections to transversality
$\mbf{k}\wedge\mbf{E} = \mbf{B}$ can be neglected, being as they are, of the
same order as neglected higher-order terms in the spatial Gaussian beam solution
to Maxwell's equations. 

We focus on the phenomenon of diffraction and specifically the detection of
photons whose wave-vectors differ significantly, either in orientation or in
magnitude from those of the background lasers. As such, we envisage an array of
photosensitive detectors being placed some distance away from the collision,
$y_{d}$, along the positive $y$-axis, much larger than the interaction volume
(the subscript $_{d}$ refers to quantities on the detector). $\mbf{E}_{b}$ is
then incident on this detector.

\section{Derivation of scattered field}
When external electromagnetic fields polarising the vacuum have equivalent
photon energies much less than the electron mass ($\hbar\omega \ll mc^{2}$),
their evolution can be well-approximated by an effective description in which
the vacuum fermion dynamics has been integrated out and only photon degrees of
freedom remain. The Euler-Heisenberg Lagrangian \cite{heisenberg_euler36} is an
effective Lagrangian which includes such fermion dynamics to one-loop order.
When the field strength is much less than critical ($E \ll E_{cr}$), the
Euler-Heisenberg Lagrangian can be well-approximated by its weak-field
expansion, which, neglecting derivative terms, is:
\be 
\mathcal{L} = \frac{1}{8\pi} (E^{2} - B^{2}) +
\frac{1}{360\pi^{2}E_{cr}^{2}} 
              \big[ (E^{2} - B^{2})^{2} + 7(\mathbf{E} \cdot \mathbf{B})^{2}
\big] + O\Big(\frac{1}{E_{cr}^{4}}(E^{2} - B^{2})\Big)^{3} +O\Big(\frac{1}{E_{cr}^{4}} \mathbf{E} \cdot
\mathbf{B}\Big)^{3}.\label{eqn:EH_Lagrangian}
\ee
\begin{figure}[!h]
\noindent\centering
\includegraphics[draft=false,
width=0.8\linewidth]{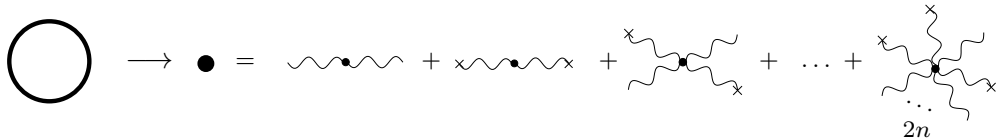}
\caption{The Euler-Heisenberg effective Lagrangian is an integration over the
high-energy (fermion) degrees of freedom. The external field can be generated by
multiple sources, indicated by photons in the diagram being with and without
crosses. \label{fig:EH1}} 
\end{figure}
The weak-field expansion \eqnref{eqn:EH_Lagrangian} is depicted in
\figref{fig:EH1} and can be understood as coupling the flux of  electromagnetic
fields from different sources with one another. Extremising
\eqnref{eqn:EH_Lagrangian} with respect to the photon gauge field returns the
wave equation for $\mbf{E}$ and $\mbf{B}$ fields modified by the one-loop,
weak-field, vacuum current, $\mbf{J}_{\trm{vac}}$:
\bea
\nabla^{2}\mbf{E} - \partial_{t}^{2}\mbf{E} &=& 4\pi\mbf{J}_{\trm{vac}}
\label{eqn:waveeqn}\\
\mbf{J}_{\trm{vac}} &=& \Big[\nabla \wedge \partial_{t}\mbf{M} -\nabla (\nabla
\cdot \mbf{P})+\partial_{t}^{2}\mbf{P}\Big], \label{eqn:jvac}
\eea
where $\mbf{P} = \frac{\partial \mathcal{L}}{\partial \mbf{E}} -
\frac{1}{4\pi}\mbf{E}$, $
\mbf{M} = \frac{\partial \mathcal{L}}{\partial \mbf{B}} + \frac{1}{4\pi}\mbf{B}$
and 
\bea
\mbf{P} & = & \phantom{-}\frac{\alpha^{2}}{180\pi^{2} m^4} \big[ 2(E^{2} -
B^{2}) \mbf{E} 
                                             + 7(\mbf{E} \cdot \mbf{B}) \mbf{B}
\big] \label{eqn:pol}\\
\mbf{M} & = & -\frac{\alpha^{2}}{180\pi^{2} m^4} \big[ 2(E^{2} - B^{2}) \mbf{B} 
                                          - 7(\mbf{E} \cdot \mbf{B}) \mbf{E}
\big]. \label{eqn:mag} 
\eea
Using the beam transversality, $\mbf{k}_{c}\wedge\mbf{E}_{c} = \mbf{B}_{c}$,
$\mbf{P}$ and $\mbf{M}$ can be written entirely in terms of the electric or
magnetic field. One can then write the vacuum polarisation as a series $P_{i} =
\chi^{(1)}_{ij}E_{j} + \chi^{(3)}_{ijkl}E_{j}E_{k}E_{l} + \ldots
\chi^{(2n+1)}_{ij\cdots m}E_{j}\cdots E_{m} + \ldots$, where electric
susceptibilities $\chi$ only occur at odd orders due to charge-conjugation
symmetry (Furry's theorem). Therefore, four-, six-, eight-, etc. wave mixing can
in principle occur, although each extra order will be suppressed by a factor
$\alpha(E/E_{cr})^{2}$. 
\newline

An iterative approach can be used to solve \eqnref{eqn:waveeqn}, which, since
$\mbf{J}^{(0)}\propto[\alpha(E^{(0)}/E_{cr})^{2}]\mbf{E}^{(0)}$ and
$\alpha(E/E_{cr})^{2}\ll1$, can be understood as perturbative:
\bea
\mbf{E}^{(n+1)}(\mbf{x}_{d},t_{d}) &=& \mbf{E}^{(n)}(\mbf{x}_{d},t_{d}) + \int
d^{3}x \frac{\mbf{J}^{(n)}(\mbf{E}^{(n)}, \ldots, \mbf{E}^{(0)}
)|_{t=t_{\trm{ret}}}}{|\mbf{x}-\mbf{x}_{d}|},\label{eqn:efield_pertexp}
\eea
where $\mbf{E}^{(n)}$ is the $n$-th order perturbative solution of
\eqnref{eqn:waveeqn}, $\mbf{E}^{(0)}$ is the zero-field vacuum solution, obeyed
by the Gaussian beams in vacuum, approximated by $\mbf{E}_{a}$ and
$\mbf{E}_{b}$, $\mbf{J}^{(n)}$ is the $n$-th iteration of the current occurring
on the right hand side of the wave equation, $(\mbf{x}_{d},t_{d})$ are the
co-ordinates in the detector plane and $t_{\trm{ret}} = t_{d}-|\mbf{x}_{d}-\mbf{x}|$ is
the retarded time. By making the approximation that:
\bea
\mbf{E}_{d}^{(n)}(\mbf{x}_{d},t_{d})=\int d^{3}x \frac{\mbf{J}^{(n)}(\mbf{E}^{(n)}, \ldots,
\mbf{E}^{(0)} )|_{t=t_{\trm{ret}}}}{|\mbf{x}-\mbf{x}_{d}|}\ll \mbf{E}^{(0)},
\label{eqn:edfield_assum}
\eea
for all $n$, the resultant electric field can be well approximated by $\mbf{E} =
\mbf{E}^{(1)} = \mbf{E}^{(0)} + \mbf{E}_{d}$, $\mbf{E}_{d} = \mbf{E}_{d}^{(0)}$,
i.e. the zero-field vacuum solution plus the lowest order ``diffracted field.''
\newline

By substituting $\mbf{E}^{(0)} = \mbf{E}_{a} + \mbf{E}_{b}$ in \eqnref{eqn:jvac}
and \eqnref{eqn:efield_pertexp}, and by enforcing the assumption that the
dimensions of the interaction volume are much smaller than the typical detector
co-ordinates, following similar steps to \cite{king10a, king10b,
dipiazza_PRL_06}, one arrives at:
\bea
\mbf{E}_{d}(\mbf{x}_{d},t_{d}) &=& -\frac{\alpha^{2}}{45\pi^{2}
m^{4}r_{d}}\frac{1}{8i} \sum_{j=1}^{12}\Big(4 +\omega_{j}^{2}\tau_{j}^{2}\Big)
\frac{A_{j}}{\tau_{j}^{2}} E_{a,0}^{B_{j}}E_{b,0}^{\Gamma_{j}} 
\mbf{v}_{l(j)}\mbox{e}^{i\psi_{j}} I_{t, j}, \label{eqn:Ed}\\
\widetilde{\mbf{E}}_{d}(\mbf{x}_{d}, \omega) &=&
-\frac{\alpha^{2}\omega^{2}}{45\pi^{3/2} m^{4}r_{d}}\frac{1}{8i} \sum_{j=1}^{12}
A_{j}\tau_{j} E_{a,0}^{B_{j}}E_{b,0}^{\Gamma_{j}}
\mbf{v}_{l(j)}\mbox{e}^{i\psi_{j}}I_{\omega, j},\label{eqn:Edtilde}
\eea
where $I_{t, j}$ and $I_{\omega, j}$ are integrals over the interaction volume,
given in \eqnrefs{eqn:It}{eqn:Iomega}, $r_{d} = |\mbf{x}_{d}|$ is the detector
distance, with $A_{j} \in \{-2, -1, 1, 2\}$, $B_{j}, \Gamma_{j} \in \{1,2\}$
coefficients given in the appendix in \tabref{tab:coeffs}, $\tau_{j} =
[B_{j}/\tau_{a}^{2} + \Gamma_{j}/\tau_{b}^{2}]^{-1/2}$, $\omega_{j} =
\beta_{j}\omega_{a} + \gamma_{j}\omega_{b}$, $\psi_{j} = \beta_{j}\psi_{a} +
\gamma_{j}\psi_{b}$, $\beta_{j}, \gamma_{j} \in \{-2, -1, 0, 1, 2\}$ also given
in \tabref{tab:coeffs} and $l(j) = 1$ for $j<=6$, otherwise $l(j)=2$ and
$\mbf{v}_{1,2}$ are the diffracted field polarisation vectors given in
\eqnrefs{eqn:v1}{eqn:v2}. 
\newline

Splitting the plane-wave part of the input fields $\mbf{E}_{a}$, $\mbf{E}_{b}$ into positive and negative frequencies, 
the twelve terms in \eqnrefs{eqn:Ed}{eqn:Edtilde} are produced, corresponding to the six possible orientations of the currents
connected by the effective vertex in \figref{fig:EH1}. As the interaction
contains terms of the order $O[(\mbf{E}_{a}+\mbf{E}_{b})^{3}]$, and as the
purely cubic terms $E_{a,b}^{3}$ necessarily disappear (both electromagnetic
invariants are zero for the individual Gaussian beams, transverse in this
approximation), for an incident current of frequency $\omega_{b}$, the resultant
signal can have a frequency $\omega_{a}, \omega_{b}, \omega_{b}\pm 2\omega_{a},
2\omega_{b}\pm\omega_{a}$, corresponding to the two beams' elastic and inelastic
components respectively. The diffracted field polarisation vectors
$\mbf{v}_{1,2}$ appear as geometrical factors and from their definition in \eqnrefs{eqn:v1}{eqn:v2},
one can see that on the detector ($y_{d}>0$), the $\omega_{a}$ and
$\omega_{a}\pm2\omega_{b}$ signal from pulse $a$ are strongly suppressed, as
would be expected as the $\mbf{E}_{a}$ pulse travels from the interaction region away from the
detector.  After a further analytical integration in $x$, the remaining
two-dimensional integrals from \eqnrefs{eqn:It}{eqn:Iomega} were then evaluated
numerically in \textsf{C++}, partly using the \textsf{GSL} library \cite{gsl}.
\newline

One can interpret the classical field incident on the detector as being composed of a total number of
photons $N_{t}$ by dividing its total energy by the photon energy so that $N_{t}
= \int^{\infty}_{-\infty} d\omega dx_{d}dz_{d}\,\widetilde{I}_{t}(\omega,
r_{d})/|\omega|$, where the total spectral density $\widetilde{I}_{t}(\omega,
r_{d}) = |\widetilde{\mbf{E}}_{t}(\omega, r_{d})|^{2}/8\pi^{2} =
|\widetilde{\mbf{E}}_{b}(\omega, r_{d})+ \widetilde{\mbf{E}}_{d}(\omega,
r_{d})|^{2}/8\pi^{2}$ ($\widetilde{\mbf{E}}_{a}(\omega,r_{d})$ is taken to be
zero in the current beam set-up) and where $y_{d}$ is taken large enough that the surface perpendicular to the Poynting vector can be well approximated as being flat. Although the spectral density extends to
negative frequencies, it is consistent to interpret the differential number of
photons as this divided by the \emph{absolute} frequency because the total
energy is the integration over all frequencies and all energy is carried by
positive-frequency photons (see also \cite{jackson75} on this point). We then
calculate the number of ``accessible'' photons that fall on the detector plane,
by integrating over the annulus that satisfies
$dN_{t}(x_{d},z_{d})-100\,dN_{b}(x_{d},z_{d})>0$ for $dN_{i} = 
\int_{-\infty}^{\infty} d\omega\,\widetilde{I}_{i}(\omega, r_{d})/|\omega|$,
$i\in\{a,b,d,t\}$. 
\section{Elastic photon-photon scattering}
\begin{figure}[!ht]
\noindent\centering
\includegraphics[draft=false,
width=0.4\linewidth]{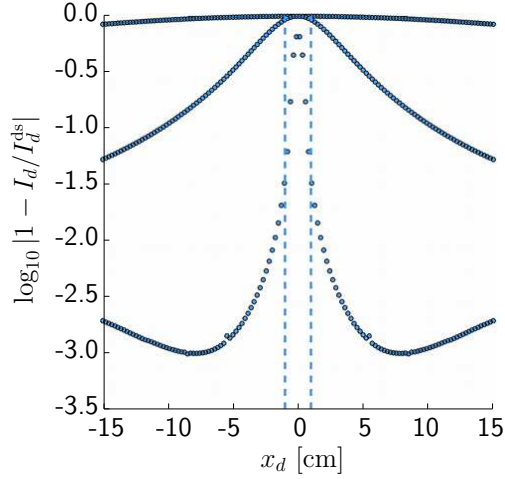}
\caption{Numerical comparison of $I_{d}$ with results in \cite{kingthesis} (denoted
$I_{d}^{\trm{ds}}$). Plotted is $\log_{10}$ of the absolute relative difference
in $I_{d}$, for, from top to bottom, $\tau_{a}=\tau_{b}=100~\trm{fs}$,
$1000~\trm{fs}$ and $20000~\trm{fs}$. Between the dotted lines, the background
dominates ($I_{b}\gg I_{d}$).\label{fig:elastic_test}} 
\end{figure}

Current and next generation high intensity lasers will typically produce pulses
with many optical cycles and so unless some resonance condition is fulfilled,
one would expect the elastic cross-section, where incident and outgoing spectra
have the same form, to be the largest. By ``elastic,'' we are therefore
referring to terms in $\mbf{E}_{d}$ with equal incoming and outgoing frequencies. As an
analytical test of our expressions, we can reproduce the electric field derived
for the three-beam, double-slit cases given in \cite{king10a, king10b} when the
separation of the slits is sent to zero - the two-beam limit (this limit was calculated for \cite{king10a} in \cite{kingthesis}), which we label
$\mbf{E}_{d}^{h}$, $\mbf{E}_{d}^{p}$ referring to head-on and perpendicular
collisions respectively. By taking $x_{0}=z_{0}=\Delta t=0$ and the limit
$\tau_{a,b}\to\infty$ in \eqnref{eqn:Ed}, with $\theta=0$, we recover
$\mbf{E}^{h}_{d}$ as given in \cite{kingthesis}, and with $\theta=\pi/2$, we recover
$\mbf{E}^{p}_{d}$ as given in \cite{king10b}. As a numerical test of our
expressions, we can compare $I_{d}(t,r_{d}) =
|\mbf{E}_{d}(t,r_{d})\cdot\mbf{E}_{d}(t,r_{d})|^{2}/4\pi$ in the case
$I_{d}(0,r_{d})$ with results using the single-slit version of $\mbf{E}_{d}^{h}$.
The equivalent parameters are $\lambda_{a}=0.8~\mu\trm{m}$,
$\lambda_{b}=0.527~\mu\trm{m}$, $w_{a,0}=0.8~\mu\trm{m}$,
$w_{b,0}=290~\mu\trm{m}$, $P_{a}=50~\trm{PW}$, $P_{b}=20~\trm{TW}$, as the field
strengths in \cite{kingthesis, king10b} were calculated using a conservative form
of the beam intensity with power per unit area for an area $\pi\trm{w}^{2}_{c,0}$,
rather than the $\pi\trm{w}^{2}_{c,0}/2$ which is manifest from an integration of
the intensity of a Gaussian beam over the transverse plane. 
In order to obtain the agreement between 
$\mbf{E}_{d}$ and $\mbf{E}_{d}^{h}$ shown in \figref{fig:elastic_test}, $\tau_{a,b}$ had to be set to around
$10^{4}~\trm{fs}$, which is unexpectedly large compared to the pulse durations
considered in those references ($\tau_{a} = 30~\trm{fs}$, $\tau_{b} =
100~\trm{fs}$). We will elaborate the non-trivial dependency of $I_{d}$ on pulse duration, which explains why most of the difference between $I_{d}$ and $I_{d}^{h}$ disappears already at $\tau = 10^{3}~\trm{fs}$. When the number of accessible photons was calculated for the
pulsed system with $\theta = 0.1$ and the same durations as suggested in
\cite{kingthesis}, the number of photons $N_{d}$ also fell from the estimated value of around
$36$ to around $0.4$. 
\newline

\begin{figure}[!ht]
\noindent\centering
\includegraphics[draft=false,
width=0.43\linewidth]{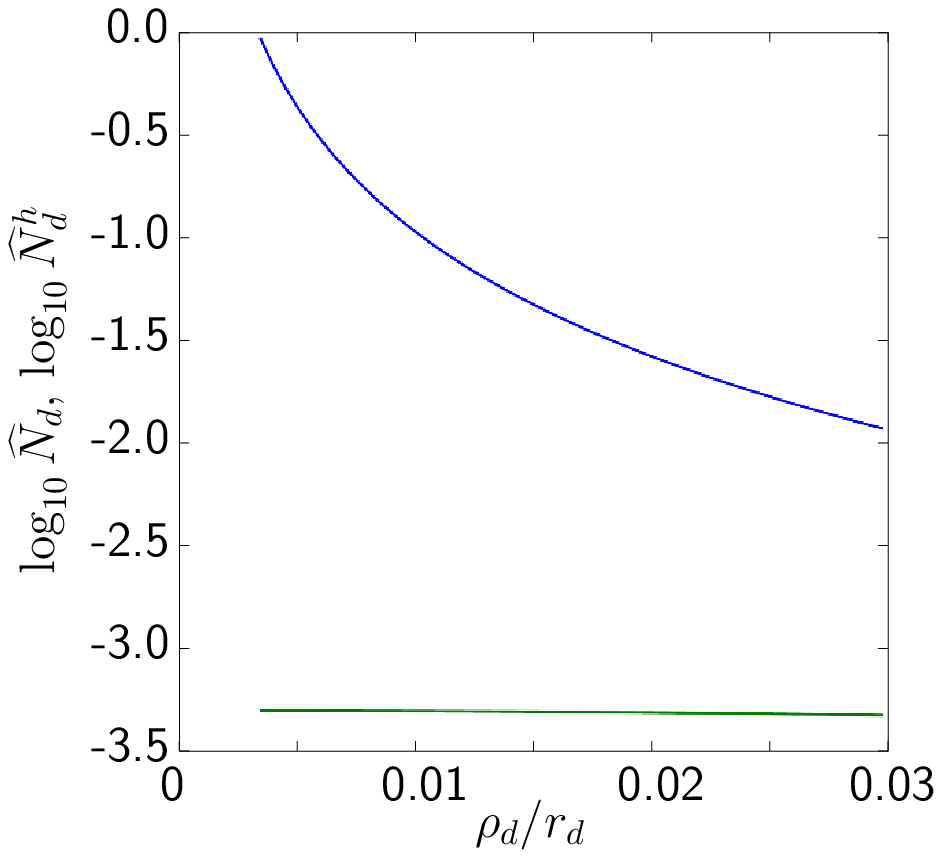}
\hspace{0.1\linewidth}
\includegraphics[draft=false,
width=0.4\linewidth]{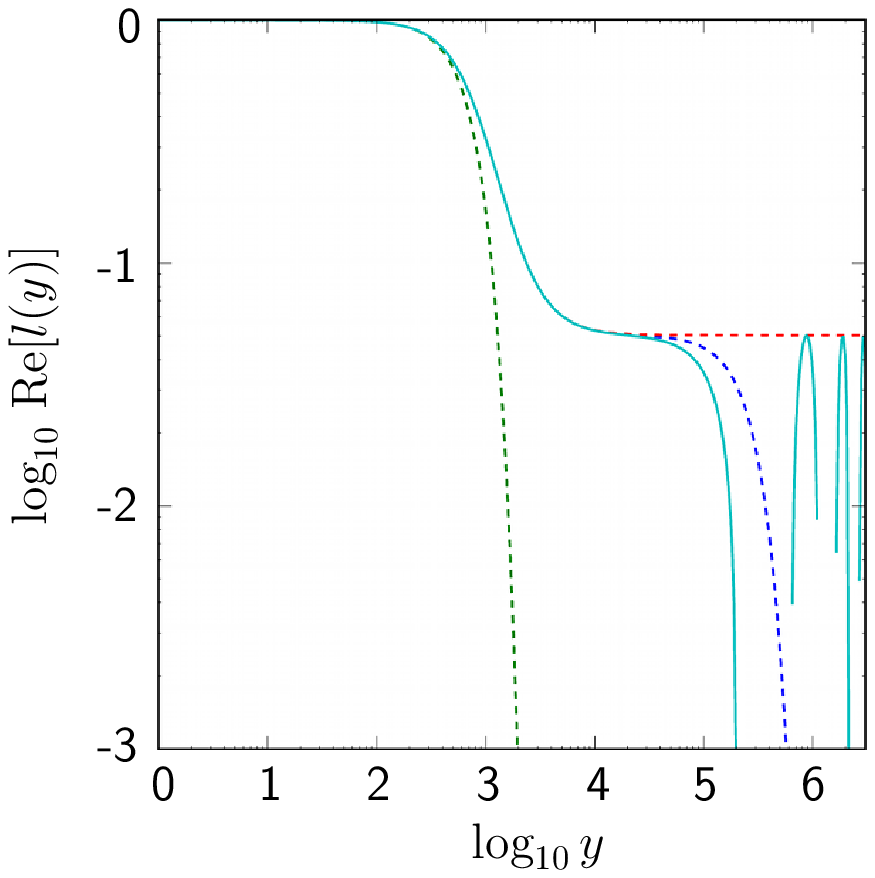}
\caption{Left-hand plot (a): $\log_{10}\widehat{N}_{d}^{h}$ (upper curve) and
$\log_{10}\widehat{N}_{d}$ (lower curve) plotted against $\rho_{d}$. Right-hand
plot (b): a log-log-plot of $\trm{Re}\,[l(y)]$ (solid line) with the first, second and third dashed lines
corresponding to $\lim_{w_{b,0}\to\infty} l(y)$, $|l(y)|$ and $\lim_{y_{r,b}\to\infty} l(y)$.
\label{fig:Nd_y_dep} }
\end{figure}

\begin{figure}[!ht]
\centering\noindent
\includegraphics[draft=false,
width=0.4\linewidth]{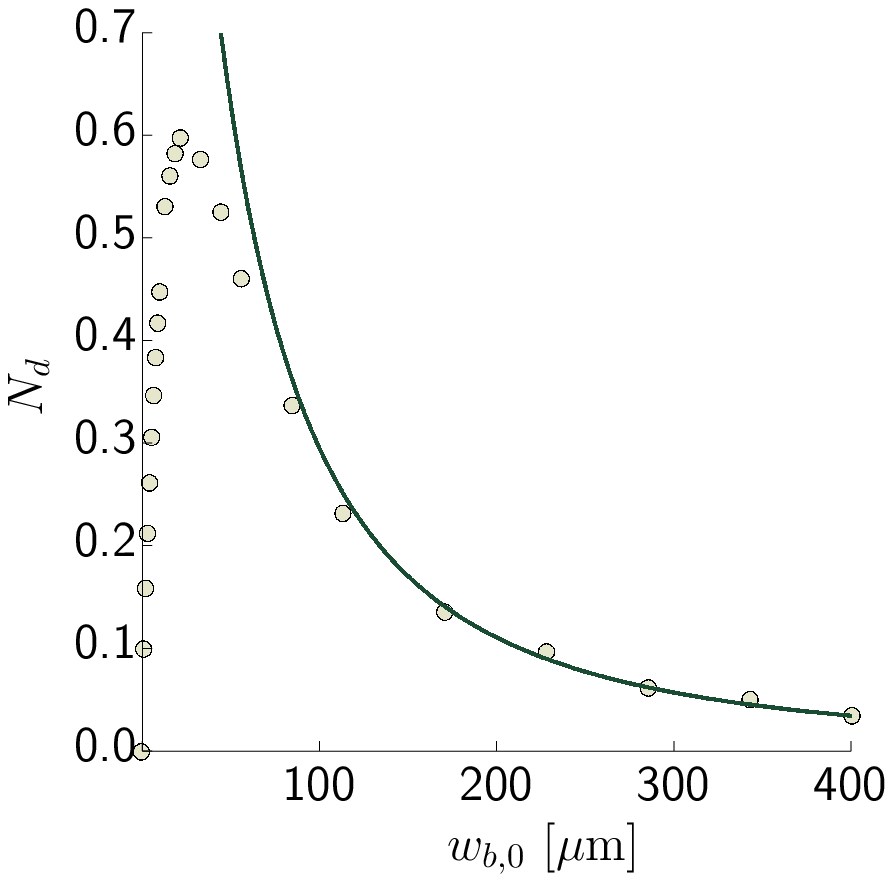}\hspace{0.1\linewidth}
\includegraphics[draft=false,
width=0.4\linewidth]{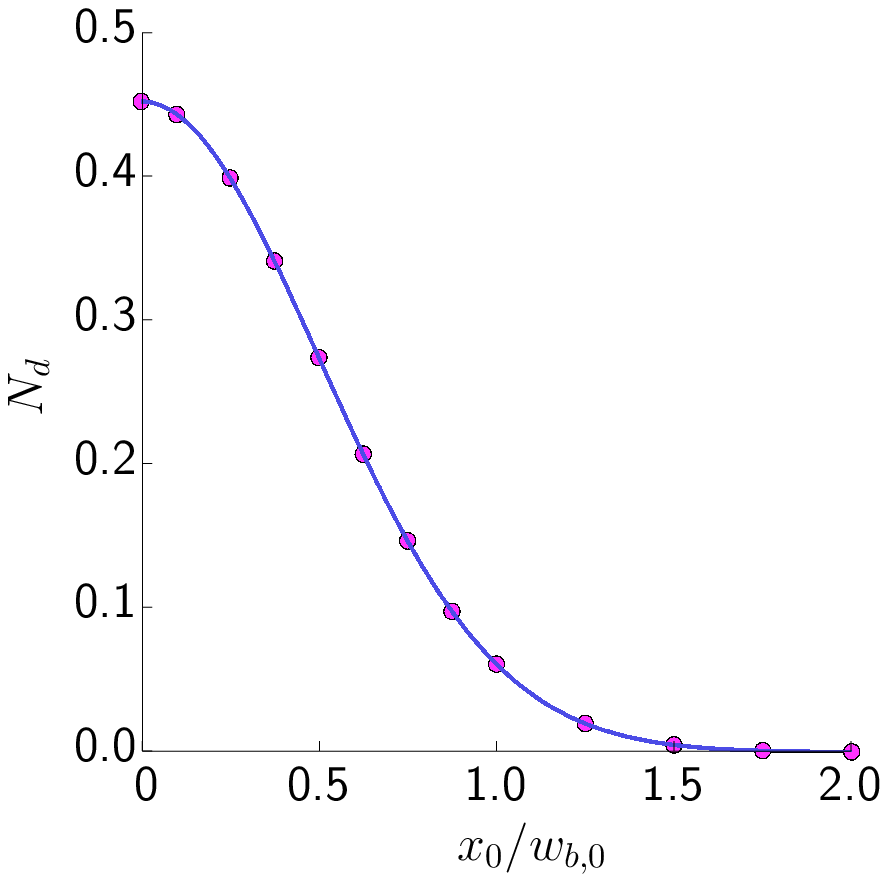}\\[2ex]
\includegraphics[draft=false,
width=0.4\linewidth]{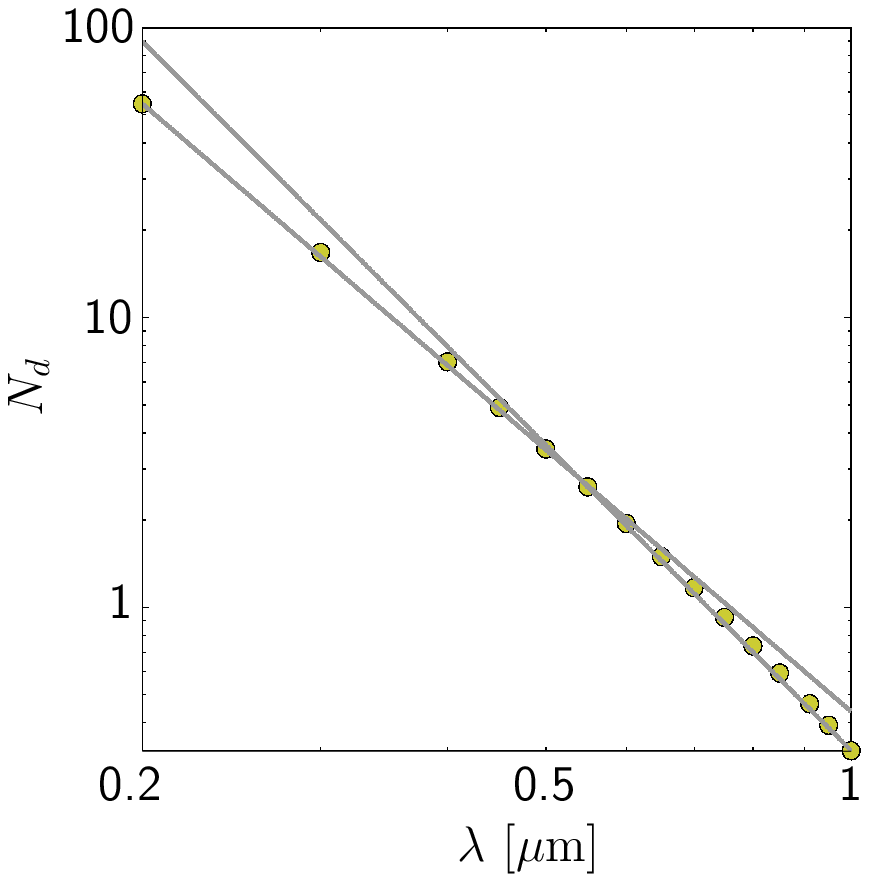}\hspace{
0.1\linewidth}
\includegraphics[draft=false,
width=0.4\linewidth]{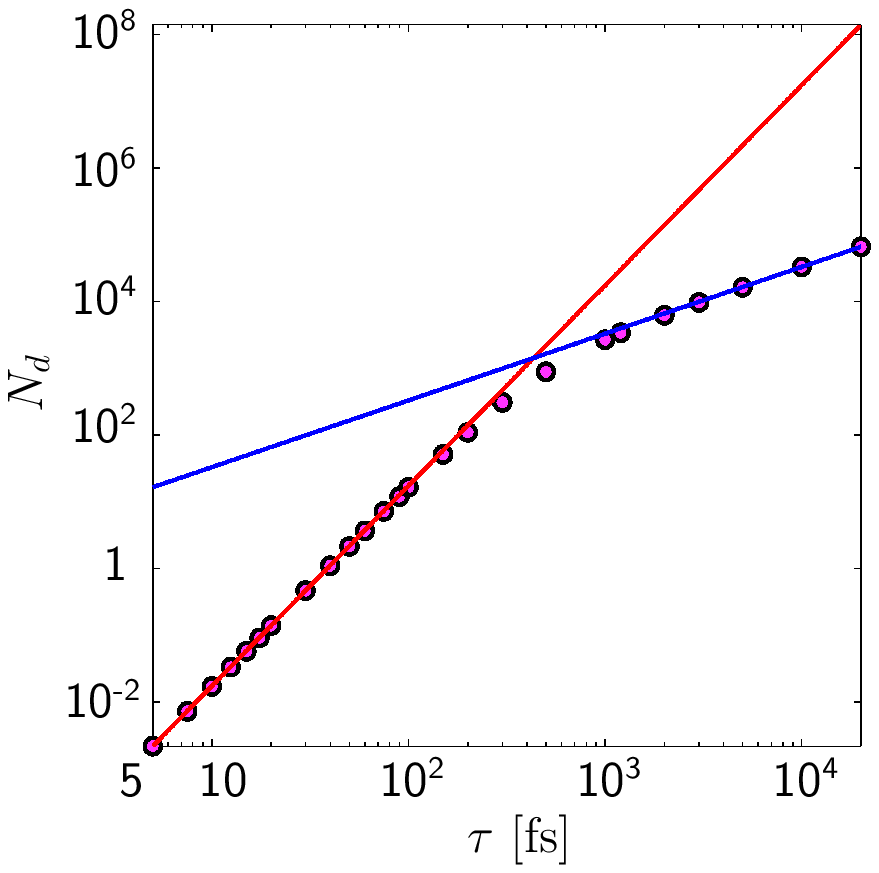}\hspace{0.1\linewidth}
\\[2ex]
\includegraphics[draft=false,
width=0.4\linewidth]{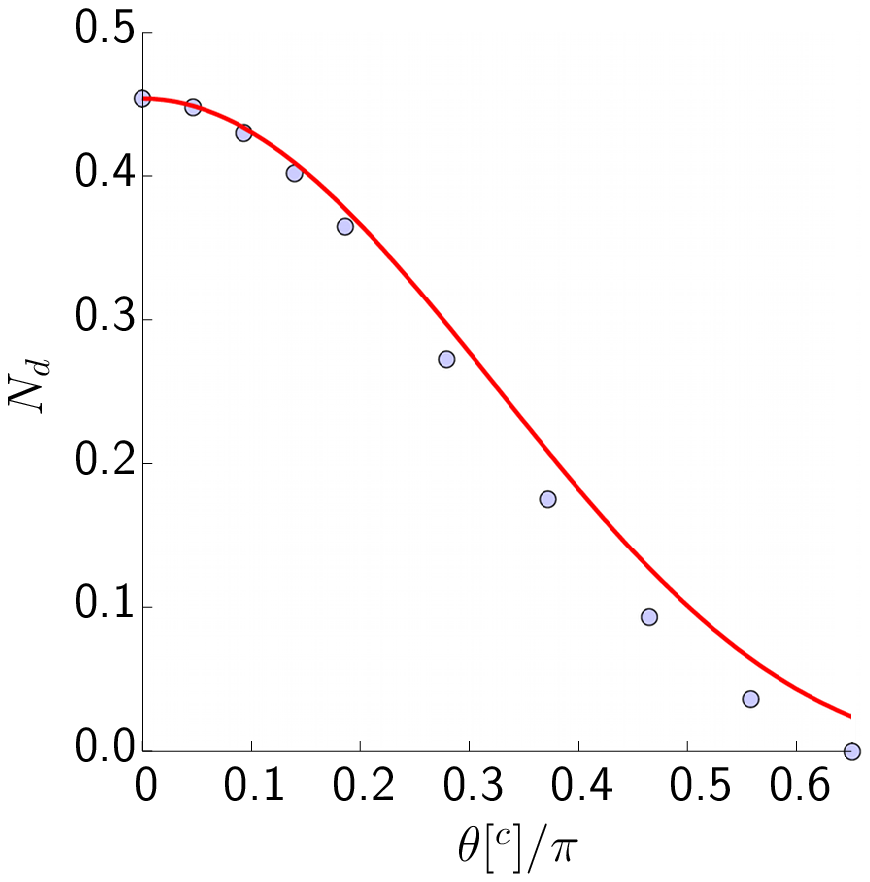}
\hspace{0.1\linewidth}
\includegraphics[draft=false,
width=0.4\linewidth]{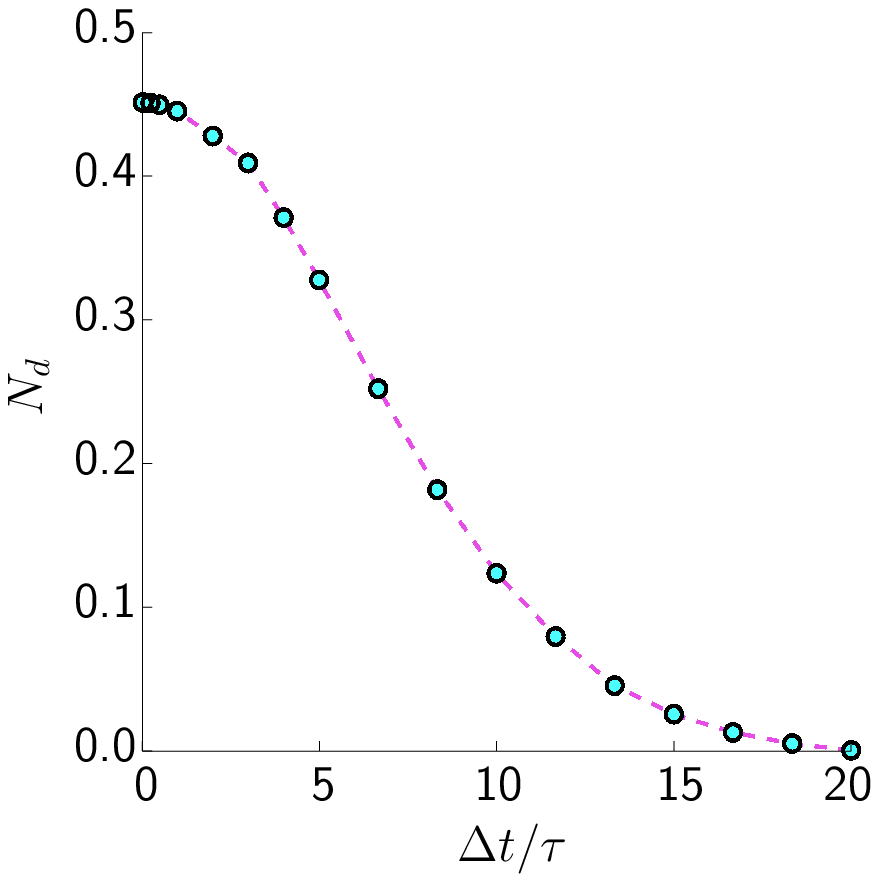}\\
\caption{Dependency of the number of measurable elastically diffracted photons
$N_{d}$ on various parameters, where parameters held constant take the values
$\lambda_{a}=\lambda_{b}=0.91~\mu\trm{m}$, $w_{a,0}=0.91~\mu\trm{m}$,
$w_{b,0}=100~\mu\trm{m}$, $\tau_{a} = \tau_{b} = 30~\trm{fs}$,
$P_{a}=5~\trm{PW}$, $P_{b}=5~\trm{PW}$, $\hat{\pmb{\eps}}_{a} =
\hat{\pmb{\eps}}_{b} = \hat{\mbf{x}}$. \label{fig:elastic_res}} 
\end{figure}

To illuminate the two orders of magnitude difference in $N_{d}$ for
these parameters, the integrand for
$\mbf{E}_{d}$ was reduced to the most significant terms for a head-on, elastic
collision and evaluated independently in \textsf{Mathematica}. The simplified
expression
$\widehat{N}_{d}(\tau)$ was then
\bea
\widehat{N}_{d}(\tau) &=&
\int^{\rho^{\trm{max}}_{d}}_{\rho^{\trm{min}}_{d}}\!\!
d\rho_{d} \rho_{d} \int_{-\infty}^{\infty} \frac{d\omega}{2\pi}
~\Big|\int_{-\infty}^{\infty}dy~ 
\frac{\sqrt{\pi}\tau}{\sqrt{3}} \mathcal{V}(y,
\omega,\omega_{b},\tau,\rho_{d}/r_{d})\Big|^{2} \label{eqn:Nd_y_dep_1}\\
\widehat{N}_{d}^{h}(\tau)&=&2\int^{\rho^{\trm{max}}_{d}}_{\rho^{\trm{
min}}_{d}}\!\!d\rho_{d}
\rho_{d}~\frac{\tau}{4}~\Big|\int_{-\infty}^{\infty}dy
\lim_{\tau\to\infty}\lim_{\omega_{b}\to\omega}
\mathcal{V}(y,\omega,\omega_{b},\tau,\rho_{d}/r_{d})\Big|^{2}\label{eqn:Nd_y_dep_2}\\
\mathcal{V}(y,\omega,\omega_{b},\tau,\rho_{d}/r_{d}) &=& \frac{\alpha
I_{a,0}E_{b,0}w_{a,0}^{2}\omega^{3/2}}{360\pi y_{d}I_{cr}i} \Big[1+\frac{1}{2}\frac{w^{2}_{a,0}}{w^{2}_{b,0}}
\frac{1+(y/y_{r,a})^{2}}{1+(y/y_{r,b})^{2}} \Big]^{-1}\nonumber\\
&&
\trm{exp}\Big[\frac{-\omega^{2}\rho_{d}^{2}(1+(y/y_{r,a})^{2})}{
4r_{d}^{2}\big[2+ \frac{w^{2}_{a,0}}{w^{2}_{b,0}}
\frac{1+(y/y_{r,a})^{2}}{1+(y/y_{r,b})^{2}}
\big]}+\frac{4i(\omega-\omega_{b})y}{3}-\frac{8y^{
2 } } { 3\tau^{2}}  -\frac{(\omega-\omega_{b})^{2}\tau^{2}}{12}
\Big], \label{eqn:Nd_y_dep_3}
\eea	
where $s^{2}=(2/w_{a,0}^{2}+1/w_{b,0}^{2})^{-1}$, $\rho_{d}^{2} =
(x_{d}^{2}+z_{d}^{2})/r_{d}^{2}$
and $\widehat{N}_{d}^{h}$ is equivalent to the expression leading from
$\mbf{E}_{d}^{h}$. The dependence on $\rho_{d}/r_d$ of these two expressions is
shown in
\figref{fig:Nd_y_dep}(a), where it can be seen that the monochromatic $\widehat{N}_{d}^{h}$ is
much larger and more sharply peaked in the
centre of the detector. 
After integrating between the relevant annulus of
$\rho_{d}^{(\trm{min})}/r_{d}=0.0032$ and
$\rho_{d}^{(\trm{max})}/r_{d}=0.03$, this independent test
then gives $\widehat{N}_{d}(\tau)=0.33$ and $\widehat{N}_{d}^{h}=37.0$. The
corresponding values for a circular detector of radius $15~\trm{cm}$ from the
full expression are $\widehat{N}_{d}(\tau)=0.33$ and $\widehat{N}_{d}^{h}=38.0$, 
supporting the two-orders of magnitude difference and the claim that \eqnreft{eqn:Nd_y_dep_1}{eqn:Nd_y_dep_3} incorporate the 
main physics. By plotting the exponential dependency on $\rho_{d}/r_d$, which is integrated over to acquire the expected number of photons 
in the monochromatic case, $\widehat{N}_{d}^{h}$:
\bea
l(y) = \trm{exp}\Big[-\frac{\omega^{2}\rho_{d}^{2}(1+(y/y_{r,a})^{2})}{
4\big[2+ \big(\frac{w_{a,0}}{w_{b,0}}\big)^ { 2 }
\frac{1+(y/y_{r,a})^{2}}{1+(y/y_{r,b})^{2}}
\big]}
\Big],
\eea
we observe the interesting behaviour shown in \figref{fig:Nd_y_dep}(b). We first note that 
the decay is not purely exponential, but has two important length scales: $w_{b,0}$ and $y_{r,b}$, which, 
in the limit of being infinitely large, correspond to the first and third dashed curves in \figref{fig:Nd_y_dep}(b).
The inclusion of these extra longitudinal length scales in \cite{kingthesis}, which cannot contribute to photon scattering 
when the finite pulse length is taken into consideration, then explains the discrepancy in the values of $\widehat{N}_{d}(\tau)$ and 
$\widehat{N}_{d}^{h}$. Only the region of the pulses within a distance $\tau$ around their maxima in the longitudinal direction can efficiently contribute to the scattering process, with the rest of the pulse being damped by its Gaussian shape.
The finite length of laser pulses probing vacuum photon-photon scattering can then
only be neglected, when the duration $\tau$ 
is the largest longitudinal length scale. In the limit $\tau\gg y_{r,b}$ in the full expression for
$\mbf{E}_{d}$ in \eqnrefs{eqn:Ed}{eqn:Edtilde}, the scaling $N_{d}(\tau) \propto
\tau$ of \cite{kingthesis} is recovered, supporting this statement (this will also be apparent from \figref{fig:elastic_res}). Furthermore, the results of \cite{kingthesis}
are expected to remain valid in the case $\pi w_{b,0}^{2}/\lambda_{b}\tau_{a,b}
< 1$, so for more focused and longer wavelength probe beams as well as for
longer pulses. Indeed for the parameters quoted, that the effect would 
be two orders of magnitude weaker is in no way prohibitive to conducting such experiments. 
For example in \cite{king10a, kingthesis}, the intensity of the probe beam
was taken to be only around $I_{p} \approx10^{16}~\trm{Wcm}^{-2}$, but as $N_d \propto I_{p}$,
the shortfall could be made up by focusing the probe beam more (if $w_{b,0}$ is
set to $60~\mu\trm{m}$, $N_d$ increases approximately by a factor $7$ in the single-slit and $4.5$ in the double-slit case) or
increasing the power of the probe (from $10~\trm{TW}$), to which $N_d$ is
proportional.   
\newline

The current treatment also allows for the two lasers to be equally strong and we
consider the more experimentally-accessible situation of having a single laser,
split into two colliding pulses, both focused to ultra-high intensities. Since
$N_{d}$ scales with $E_{a,0}^{2B_{j}}E_{b,0}^{2\Gamma_{j}}$ if we keep
the power of the laser constant ($E_{c,0} = 2\sqrt{2 P_{c,0}}/w_{c,0}$), for
each term, the optimal division of the total power between the
beams is: 
\bea
\frac{P_{a,0}}{P_{b,0}} = \frac{B_{j}}{\Gamma_{j}}; \qquad P_{a,0} =
\frac{B_{j}}{B_{j}+\Gamma_{j}}P_{t,0}.
\eea
For base parameters similar to that of the Vulcan laser \cite{vulcan_site}
$\lambda_{a}=\lambda_{b}=0.91~\mu\trm{m}$, $\tau_{a} = \tau_{b} = 30~\trm{fs}$,
$P_{a}=5~\trm{PW}$, $P_{b}=5~\trm{PW}$, with $w_{a,0}=0.91~\mu\trm{m}$,
$w_{b,0}=100~\mu\trm{m}$, $\hat{\pmb{\eps}}_{a} = \hat{\pmb{\eps}}_{b} =
\hat{\mbf{x}}$ a summary of the dependency of $N_{d}$ on several variables is
given in \figref{fig:elastic_res}. We will comment on the plots sequentially, in
which solid lines represent what one could intuitively expect, as explained in
the following. Starting from the right-hand side of the first plot and moving in
the direction of falling $w_{b,0}$, we see  $N_{d}(w_{b,0})$ increases
approximately as $\propto w_{b,0}^{-2}$, indicated by the solid line. Since
$N_{d}$ for such a set-up is proportional to $E_{b,0}^{2}$, and since
this is inversely proportional to the area of focusing, the dependency on
$\propto w_{b,0}^{-2}$ is as expected. Deviation occurs when a maximum is
reached (see e.g. \cite{king10b} for details), beyond which $N_{d}(w_{b,0})$
falls rapidly as the background from $\mbf{E}_{b}$ gradually covers the entire
detector, leaving no signal. The dependency on beam-separation $N_{d}(x_{0})$ is
also intuitive and seen to have excellent agreement with a Gaussian, normalised
in height, with a width of $w_{b,0}/2$ ($\exp(-2x_{0}^{2}/w_{b,0}^{2})$). Simply
by integrating the transverse Gaussian distributions of the two beams, and
then squaring ($N_{d} \propto |\mbf{E}_{d}|^{2}$), one arrives at this
dependency. The third plot of $N_{d}(\lambda)$ ($\lambda = \lambda_{a} =
\lambda_{b}$) is a log-log plot where the dependency begins for small $\lambda$
as $N_{d}\approx \lambda^{-3}$ but then for larger values tends to
$N_{d}\approx\lambda^{-3.5}$.  This is shown by all the points lying between
these two solid lines. Since the power of each beam is inversely proportional to
wavelength, and since the $N_{d} \propto P_{a,0}^{2}P_{b,0}$, one would expect
at least a dependency of $N_{d}(\lambda) \sim \lambda^{-3}$. In contrast, the dependency of
$N_{d}(\tau)$ can be straightforwardly derived. For $\tau_{a} = \tau_{b} =
\tau$, one notes that when $\tau \ll w_{b,0}, y_{r,b}$, the interaction volume in beam
propagation direction is governed by the Gaussian pulse shape. Further noting
that $N_{d}$ essentially involves a double integration on longitudinal beam
co-ordinate (through taking the mod-squared), as well as an integral over $t$,
the dependency $N_{d}(\tau)\propto \tau^{3}$ appears, which shows excellent
agreement for small $\tau$ with the full numerical integration, displayed by the on the log-log plot of
$N_{d}(\tau)$ in the fourth figure. The larger $\tau$ is for $\tau > w_{b,0}$, the more the decay along the beam propagation axis is described by focusing rather than pulse terms. For large enough $\tau$, $I_{d}$ depends only on focusing terms and since the yield $N_{d}$ is acquired from an integration over time, we have $N_{d}(\tau) \propto \tau$ and this transition can be seen by the second, linear, fit line for large $\tau$ in the figure.  An estimation of the dependency of
$N_{d}(\theta)\propto(1+\cos\theta)^{2}$ on beam intersection angle comes from
the geometrical factor in $\mbf{v}_{1} \propto (1+\cos\theta)$, which must be
squared and gives the approximate agreement shown in the fifth plot. For small
angles $N_{d}(\theta) \propto 1- \theta^{2}/2$, making the dependency relatively
weak for near head-on collisions ($N_{d}$ remains at $90\,\%$ of its value up
to $\theta \approx \pi/7$). The final plot of $N_{d}(\Delta t)$ closely resembles a Gaussian
with width $9\tau$ and so for this set-up, $N_{d}$ is relatively insensitive to
lag.
\newline

One strategy to increase the number of diffracted photons would be to use
higher-harmonics of the probe laser. If the same parameters as in
\figref{fig:EH1} are used, for a collision angle of $\theta = 0.1$, assuming a
$40\%$ reduction in energy due to generating the second harmonic, $N_{d} \approx
4$. If this process could be repeated to generate the fourth-harmonic, with a
$16\%$ reduction, $N_{d} \approx 13$. As previously argued in \cite{king10a},
such numbers of scattered photons should allow detection in experiment. A
discussion of sources of background noise and why they can be effectively
neglected is given in \cite{king10a, kingthesis}. 



\section{Inelastic photon-photon scattering (four-wave mixing)}

When considering the possible frequencies of the resultant current, conservation
of energy and linear momentum leads one to the equations:
\bea
\omega &=& \label{eqn:4wave_energy_cons} \trm{sgn}(\beta_{j})[\omega_{a,1} +
\delta_{|\beta_{j}|2}\omega_{a,2} ] +
\trm{sgn}(\gamma_{j})[\omega_{b,1} + \delta_{|\gamma_{j}|2}\omega_{b,2} ]\label{eqn:4wave_mom_cons0}\\
\omega\, \frac{y_{d}}{r_{d}}  &=&
\trm{sgn}(\beta_{j})[\omega_{a,1}\cos\theta_{a,1} +
\delta_{|\beta_{j}|2}\omega_{a,2}\cos\theta_{a,2} ] \nonumber \\ &&
+~\trm{sgn}(\gamma_{j})[\omega_{b,1}\cos\theta_{b,1} +
\delta_{|\gamma_{j}|2}\omega_{b,2}\cos\theta_{b,2} ] \label{eqn:4wave_mom_cons1}
\\
\omega\, \frac{\rho_{d}}{r_{d}} &=&
\trm{sgn}(\beta_{j})[\omega_{a,1}\sin\theta_{a,1} +
\delta_{|\beta_{j}|2}\omega_{a,2}\sin\theta_{a,2}] \nonumber \\
&&+~\trm{sgn}(\gamma_{j})[\omega_{b,1}\sin\theta_{b,1} +
\delta_{|\gamma_{j}|2}\omega_{b,2}\sin\theta_{b,2}] \label{eqn:4wave_mom_cons2},
\eea
where $\omega$ is the frequency of the resultant current, $\trm{sgn}(x)$ returns
the sign of $x$ with $\trm{sgn}(0)=0$, and $\theta_{\{a,b\},\{1,2\}}$ are the
angles the currents make with $\hat{\mbf{y}}$. Therefore, detection co-ordinate,
focusing and harmonic order are already linked at this stage. It turns out to be
difficult to satisfy these conditions simultaneously with just two laser beams and a fixed observation angle.
For example, if we take $\beta_{j} = 2$, $\gamma_{j} =1$, with $\omega_{a,1} = \omega_{a,2} = \omega_{a}$, $\omega_{b,1} = \omega_{b,2} = \omega_{b}$ for simplicity and a more-or-less head-on collision of the lasers, so $\theta_{a,\{1,2\}}$ is approximately 
equal to $\pi - \theta_{b,\{1,2\}}$ and $\theta_{b,\{1,2\}}$ is small, then, to first order,
from \eqnrefs{eqn:4wave_mom_cons0}{eqn:4wave_mom_cons1} we have $\omega = 2\omega_{a} +\omega_{b}$ and $\omega\,y_{d}/r_{d} \approx -2\omega_{a}+\omega_{b}$. Since $y_{d}/r_{d}\approx 1$ on the detector, the contribution from this term can therefore only be satisfied by a small range of frequencies around $\omega_{a} = 0$, which are not typically populated in the spectrum of $\mbf{E}_{a}$. The energy-momentum
conditions \eqnreft{eqn:4wave_energy_cons}{eqn:4wave_mom_cons2} can be most
easily seen occurring in the exponent of the integral $I_{\omega}$
\eqnref{eqn:Iomega}, where they appear as frequencies of plane waves to be
integrated over in $y$, $z$, becoming Gaussian-like after integration.
The larger the deviation from these conditions, the higher the
frequency of oscillation to be integrated over, the more exponentially small the
resulting amplitude, typical for evanescent waves.
\newline

We investigated the ansatz that for short enough pulses, the bandwidth of the
two lasers becomes wide enough that
\eqnreft{eqn:4wave_energy_cons}{eqn:4wave_mom_cons2} can be fulfilled
simultaneously for a measurable amount of photons. Essentially, for this
four-wave interaction, three different photon energies can be supplied by two
lasers. To make this statement explicit, instead of using a temporal envelope,
we can consider building the pulses in the frequency domain:
\bea
\mbf{E}'_{c}(\mbf{x}, t) = \int^{\infty}_{-\infty} d\omega_{c}\,
\mbf{E}^{\trm{mono}}_{c}(\mbf{x},t, \omega_{c}) g(\omega_{c},\omega_{c,0}),
\eea
where $\mbf{E}^{\trm{mono}}_{c}(\mbf{x},t, \omega_{c})$ is the electric field of
a monochromatic Gaussian beam, frequency $\omega_{c}$ and $g(\omega_{c},
\omega_{c,0})$ is the spectral density of the pulse $\mbf{E}'_{c}(\mbf{x}, t)$,
with peak frequency $\omega_{c,0}$. Then due to our interaction being cubic in
the fields ($\mbf{E}_{a}^{B_{j}}$, $\mbf{E}_{b}^{\Gamma_{j}}$), the integration
over this current in the frequency domain, \eqnref{eqn:Iomega}, would include
three extra integrations over frequency $\int
d\omega_{a}d\omega_{b}d\omega_{c}\,g(\omega_{a},\omega_{a,0})\,g(\omega_{b},
\omega_{b,0})\, g(\omega_{c},\delta_{B_{j}2}\omega_{a,0} +
\delta_{\Gamma_{j}2}\omega_{b,0})\,\delta(\omega - \omega_{j})$, where the final
delta function appears explicitly from an integration over $t$. Here it is
apparent that due to the finite bandwidth, in general, three different energies
enter the effective vertex in \figref{fig:EH1} from the two lasers. If the
spectrum is taken to be Gaussian $g(\omega_{c}, \omega_{c,0}) =
\exp[-(\omega_{c}-\omega_{c,0})^{2}\tau_{c}^{2}/4]\tau_{c}/2\sqrt{\pi}$ we have, setting $\theta=0$
without loss of generality:
\bea
\mbf{E}'_{c}(\mbf{x}, t) &=& \hat{\pmb{\eps}}_{c}\int_{-\infty}^{\infty}\!\! d\omega_{c}
\frac{1}{2i}\frac{E_{c,0}\,\mbox{e}^{-\frac{x^{2}+z^{2}}{w_{c}^{2}(y)}-(\omega_{
c}-\omega_{c,0})^{2}\frac{\tau_{c}^{2}}{4}}}{\sqrt{1+(y/y_{r,c})^{2}}}\mbox{e}^{i(\omega_{
c}(y-t) + \tan^{-1}\frac{2y}{\omega_{c} w_{c,0}^{2}} -
\frac{2\omega_{c}y(x^{2}+z^{2})}{4y^{2}+\omega_{c}^{2}w_{c,0}^{4}y^{2}} )} 
 + \trm{c. c.}.\nonumber\\
& = &
\frac{E_{c,0}\,\mbox{e}^{-\frac{x^{2}+z^{2}}{w_{c}^{2}(y)}-(\frac{y-t}{\tau_{c}}
)^{2}}}{\sqrt{1+(y/y_{r,c})^{2}}}\sin\Big[\omega_{c,0}(y-t) +
\tan^{-1}\frac{y}{y_{r}} -
\frac{\omega_{c,0}y}{2}\frac{x^{2}+z^{2}}{y^{2}+y_{r}^{2}} \Big] + \trm{h. o.
t.}, \nonumber
\eea
where the remaining terms are of the same order as those neglected in the
Gaussian beam solution. Therefore the use of a Gaussian temporal envelope in
$\mbf{E}_{a}$ and $\mbf{E}_{b}$ (\eqnrefs{eqn:Ea}{eqn:Eb}), is equivalent to
integrating over three different photon frequencies from the external fields in
the interaction.
\newline

\begin{figure}[!ht]
\noindent\centering
\includegraphics[draft=false,
width=0.4\linewidth]{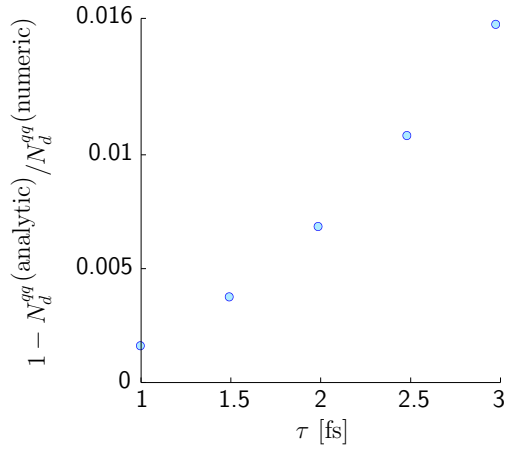}
\caption{Agreement of the analytical approximation \eqnref{eqn:dNqq_analytic}
with the corresponding numerical solution. \label{fig:dN_agreement}} 
\end{figure}

When $x_{0} = z_{0} = \Delta t = \theta = 0$, $y_{r} = y_{r,a} = y_{r,b}$ and
$\rho_{d}^{2}/r_{d}^{2} = (x_{d}^{2}+z_{d}^{2})/r_{d}^{2}$ is small,
$dN_{d}(x_{d},z_{d})$ can be approximated analytically. We can write
$dN_{d}(x_{d},z_{d}) = \sum_{p,q=1}^{12} dN^{pq}_{d}(x_{d},z_{d})$ and
demonstrate this analysis by concentrating on a single term $N_{d}^{qq}$ for
convenience (the full expression is given in \eqnref{eqn:dN_analytic}). One can show:
\bea
dN^{qq}_{d}(x_{d},z_{d}) &\approx& \frac{2}{\pi^{2}} \Bigg[\frac{\alpha
A_{q}}{90}\frac{E_{a,0}^{B_{q}}E_{b,0}^{\Gamma_{q}}}{E_{cr}^{2}}\frac{|\mbf{v}_{
l(q)}|}{16}\frac{w_{a,0}^{2}}{sr_{d}}\frac{\tau_{q}^{2}}{\sqrt{1-\tau_{qq}^{4}}}
\Bigg]^{2}  \mbox{e}^{-\frac{\omega_{q}^{2}\tau^{2}}{2}[1 +
\frac{(\tilde{\omega}_{qq} - \tau_{qq}^{2})^{2}}{1 - \tau_{qq}^{4}}]} \nonumber
\\
&& \int^{\infty}_{-\infty} d\omega |\omega|^{3}
\mbox{e}^{-[\frac{\rho_{d}^{2}\trm{w}^{2}_{a,0}}{2sr_{d}^{2}} +
\frac{\tau_{q}^{2}}{2} + \frac{\tau_{q}^{2}(\tau_{qq}^{2} +
y_{d}/r_{d})^{2}}{2(1-\tau_{qq}^{4})}]\omega^{2} + \tau_{q}^{2}\omega_{q}[1 -
\frac{(y_{d}/r_{d} + \tau_{qq}^{2})(\tilde{\omega}_{qq} - \tau_{qq}^{2})}{1 -
\tau_{qq}^{4}}]\omega},\label{eqn:Nd_analytic_y}
\eea
where $s = 1/(B_{j} + \Gamma_{j}(w_{a,0}/w_{b,0})^{2})$,  $\tau_{qq} = \tau_{q}/\ttau_{q}$, 
$\tilde{\tau}_{q}^{2} = (B_{q}/\tau_{a}^{2}-\Gamma_{q}/\tau_{b}^{2})^{-1}$ and
$\tilde{\omega}_{qq} = \tilde{\omega}_{q}/\omega_{q}$, $\tilde{\omega}_{q} =
\beta_{q}\omega_{a} - \gamma_{q}\omega_{b}$, under the condition 
$T^2/y_r^2 \ll 1$, for $T^{2}=\tau_{q}^{2}[1-\tau_{qq}^{ s4}]^{-1}$ and where a condition on $\omega$:
$|[\omega (\tau_{qq}-y_{d}/r_{d}) + \tilde{\omega}_{q}-\omega_{q}\tau_{qq}^{2}]T^{2}/y_{r}|\ll 1 $ has been approximated by taking the upper limit of the integration as $\infty$. To simplify the discussion, let $\tau_{a} = \tau_{b} = \tau$. Then we can see from
\eqnref{eqn:Nd_analytic_y} that the spectral density for inelastically scattered
photons has a different shape to the background, namely with a minimum at
$\omega = 0$ and two maxima, whose positions for the case $x_{d}=z_{d}=0$ are
$\omega^{\pm} =
(\gamma_{q}\omega_{b}/2)(1\pm[1+12/(\gamma_{q}\omega_{b}\tau)^{2}]^{1/2})$.
Using a spectral filter, and short enough pulses, this could in principle be
used to separate the different inelastic scattering signals from each other and
the elastically scattered and background photons on the detector. Setting
$\rho_{d}=0$ for brevity, the final integral can be approximated by:
\bea
dN^{qq}_{d}(x_{d},z_{d}) &=& \sqrt{\frac{2}{\pi^{3}}} \Bigg[\frac{\alpha
A_{q}}{180}\frac{E_{a,0}^{B_{q}}E_{b,0}^{\Gamma_{q}}}{E_{cr}^{2}}\frac{|\mbf{v}_
{l(q)}|}{16}\frac{w_{a,0}^{2}}{sr_{d}}\Bigg]^{2}
\big(\gamma_{q}\omega_{b}\tau\big)[3 + (\gamma_{q}\omega_{b}\tau)^{2}]
\nonumber\\ &&
\qquad\qquad\qquad\qquad\qquad\qquad\qquad\trm{Erf}\Bigg[\frac{\gamma_{q}\omega_
{b}\tau}{\sqrt{2}}\Bigg]\mbox{e}^{-\frac{1}{4}(\beta_{q}\omega_{a}\tau)^{2} -
\frac{1}{18}(\gamma_{q}\omega_{b}\tau)^{2}}
 \label{eqn:dNqq_analytic}
\eea
It should be noted that $\mbf{v}_{l(j)}$ is identically zero for $j>6$ at $r_{d}
= y_{d}$, and so the frequencies $\omega_{a}$, $2\omega_{b}\pm \omega_{a}$ are
suppressed, as already argued.  The numerical integration of the full
highly-oscillating integrands was performed using the Filon method, which is an
approximation to the integral $\int dt f(t) \cos(\omega t)$ for
asymptotically-large $\omega$ (see e.g. \cite{rabinowitz67}), used with the
\textsf{GNU} arbitrary-precision \textsf{C++} library \cite{MPC}. Agreement
between numerics and analytics for $w_{a,0}=w_{b,0}=10~\mu\trm{m}$,
$y_{d}=1~\trm{m}$, $\beta_{q}=2$, $\gamma_{q}=1$ is then shown in
\figref{fig:dN_agreement}, in part corroborating our numerical approach.
\newline

\begin{figure}[!h]
\noindent\centering
\includegraphics[draft=false,
width=0.4\linewidth]{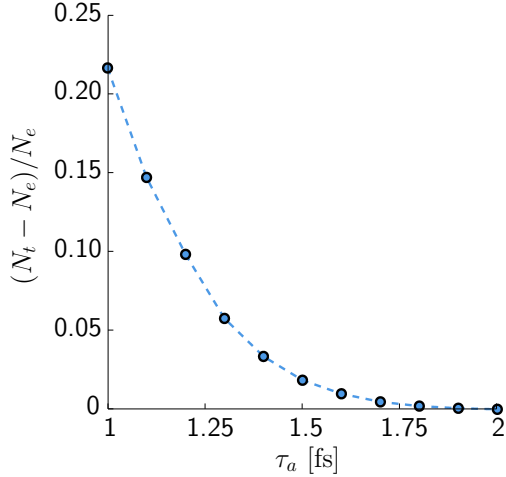}
\caption{The increasing importance of the inelastic process with increasing
bandwidth. Plotted is the proportion of the total number of diffracted photons
that are due to inelastic scattering, against $\tau_{a}$, for
$P_{b}=10/3~\trm{PW}$, $P_{a}+P_{b} = 10~\trm{PW}$, $\lambda_{a} =
0.91~\mu\trm{m}$, $\lambda_{b} = 0.2275~\mu\trm{m}$, $\tau_{b} = 2~\trm{fs}$,
$w_{a,0} = 0.91~\mu\trm{m}$, $w_{b,0} = 50~\mu\trm{m}$,
$\pmb{\eps}_{a}=(1,0,0)$, $\pmb{\eps}_{b}=(0,0,1)$, $\psi_{a} = \psi_{b} = 0$. }
\label{fig:IE1} 
\end{figure}

The pulse duration of each laser plays an important role in four-wave mixing. By
choosing a temporal profile for the beam that is Gaussian, we already have implicitly
the lower bound $\tau \gg 1/\omega$. As pulse duration and longitudinal
co-ordinate are linked, a natural upper bound is also formed for our calculation
in the assumption that the diffracted field is smaller than the
vacuum-polarising fields \eqnref{eqn:edfield_assum}. Assuming scattered photons 
arriving at a point on the detector are generated in the centre of the beams' 
intersection, the integration is exclusively over regions in which the polarising beams
are more intense than the diffracted field when $\tau \ll 2 w_{b,0}
y_{d}/\rho_{d}$, giving $1/\omega \ll \tau \ll 2 w_{b,0} y_{d}/\rho_{d}$. The
lower bound limits our ability to assess the importance of the inelastic
process. We require a large bandwidth $\Delta \omega/\omega$ for the
inelastically-generated photons to be on-shell, but from the bandwidth theorem,
$\Delta\omega/\omega \sim 1/\omega\tau \ll 1 $ by our limitation on $\tau$. As a
consequence, with a two-beam set-up, spectrally separating off the inelastic
signal would be experimentally challenging, as this signal is generated when the
bandwidth of the elastic background overlaps these ``inelastic'' frequencies.
More promising seems to be to observe the change in $N_{d}$ due to inelastic
scattering becoming significant as $\tau$ is reduced. In \figref{fig:IE1}, we
plot this ratio $(N_{t}-N_{e})/N_{e}$ against $\tau_{a}$, where $N_{e}$ is the
number of photons scattered due to when only the elastic terms are included in
\eqnref{eqn:Edtilde}. The results suggest that for short enough pulse durations,
the inelastic process can influence the total number of measured photons
substantially. In \figref{fig:IE1} the proportion reaches over 20\%, for a
minimum pulse duration of $\tau = 1~\trm{fs}$, equivalent to $\omega_{a}\tau_{a}
\approx 2$. This could already have been anticipated from
$\mbf{E}_{d}(\mbf{x}_{d}, t_{d})$ in \eqnref{eqn:Ed}, including, as it does, a
pre-factor $4 + (\omega_{j}\tau_{j})^{2}$. In addition, although the pulse
durations are short, assuming again $40\%$ attenuation each time a
second-harmonic is generated from the probe, the total number of diffracted
photons ranges from $1$ to $4$ (at $\tau_{a}=1$, $2$ respectively).  Although
the analysis is limited by how small $\tau_{a}$ can be consistently made, these
results lend support to the ansatz that two laser beams with a large bandwith,
especially in the laser being probed, can be used to measure the effect of the
inelastic process.
\newline

In order to further support this ansatz and without being limited by a minimum
value of the pulse duration, we can consider the simplified case of the
collision of two plane waves modulated by a $\trm{sech}$ envelope.
\bea
\mbf{E}_{a}(y+t) &=& \hat{\pmb{\varepsilon}}_{a}\, E_{a,0}
\cos(\omega_{a}(y+t))\,\trm{sech}\Big[\frac{y+t}{\tau_{a}}\Big]\\
\mbf{E}_{b}(y-t) &=& \hat{\pmb{\varepsilon}}_{b}\, E_{b,0}
\cos(\omega_{b}(y-t))\,\trm{sech}\Big[\frac{y-t}{\tau_{b}}\Big].
\eea
These fields satisfy Maxwell's vacuum equations exactly, therefore removing the
limitations on conceivable pulse lengths brought about by using a perturbative
solution. The analysis proceeds just as for the Gaussian case but with the
difference that now the fields are not bound in the transverse plane. Therefore,
in order to avoid a divergence, we only consider the resulting $\mbf{P}$ and
$\mbf{M}$ to be non-zero up to a finite transverse radius $\rho_{0}$. It can be
shown that this curtailing of the interaction region then allows us to integrate
over the current \eqnref{eqn:efield_pertexp} as usual. The diffracted field
$\widetilde{\mbf{E}}^{\trm{sech}}_{d}(\mbf{x}_{d}, \omega)$ then becomes:
\bea
\widetilde{\mbf{E}}^{\trm{sech}}_{d}(\mbf{x}_{d},\omega) &=&
-\frac{\omega^{2}\alpha^{2}}{45\pi^{2}m^{4}r_{d}}\frac{1}{8i} 
\sum_{j=1}^{2}E_{a,0}^{3-j}E_{b,0}^{j}\mbf{v}_{j}
I^{\trm{sech}}_{\omega,j},\label{eqn:Edsech}
\eea
where $\mbf{v}_{j}$ are geometrical factors as in the Gaussian case
\eqnrefs{eqn:v1}{eqn:v2}, $I^{\trm{sech}}_{\omega,j}$ are integrals given in
\eqnref{eqn:Isech}, the sum over $j$ corresponds to the two terms 
$E_{a,0}^{2}E_{b,0}$ and $E_{a,0}E_{b,0}^{2}$ respectively and $z_{d}=0$ has
been set for simplicity. Unlike for Gaussian beams, the elastic scattering terms
cannot be isolated so easily. In order to exemplify the effect of the inelastic
process however, one can observe how the behaviour of $N^{(\trm{sech})}_{d}$
changes as $\omega_{a}\tau_{a}$ is reduced to below unity. Deviation from
``elastic'' behaviour, indicates the importance of inelastic scattering.
\newline 

\begin{figure}[!h]
\noindent\centering
\includegraphics[draft=false,
width=0.4\linewidth]{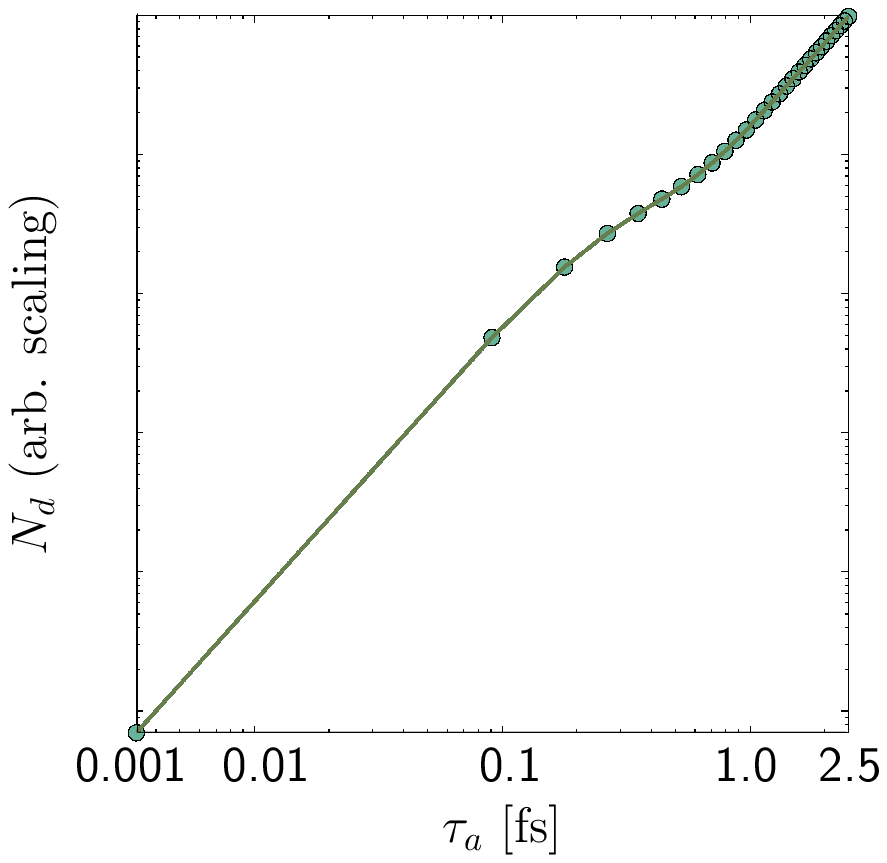}\hspace{
0.07\linewidth}
\includegraphics[draft=false,
width=0.43\linewidth]{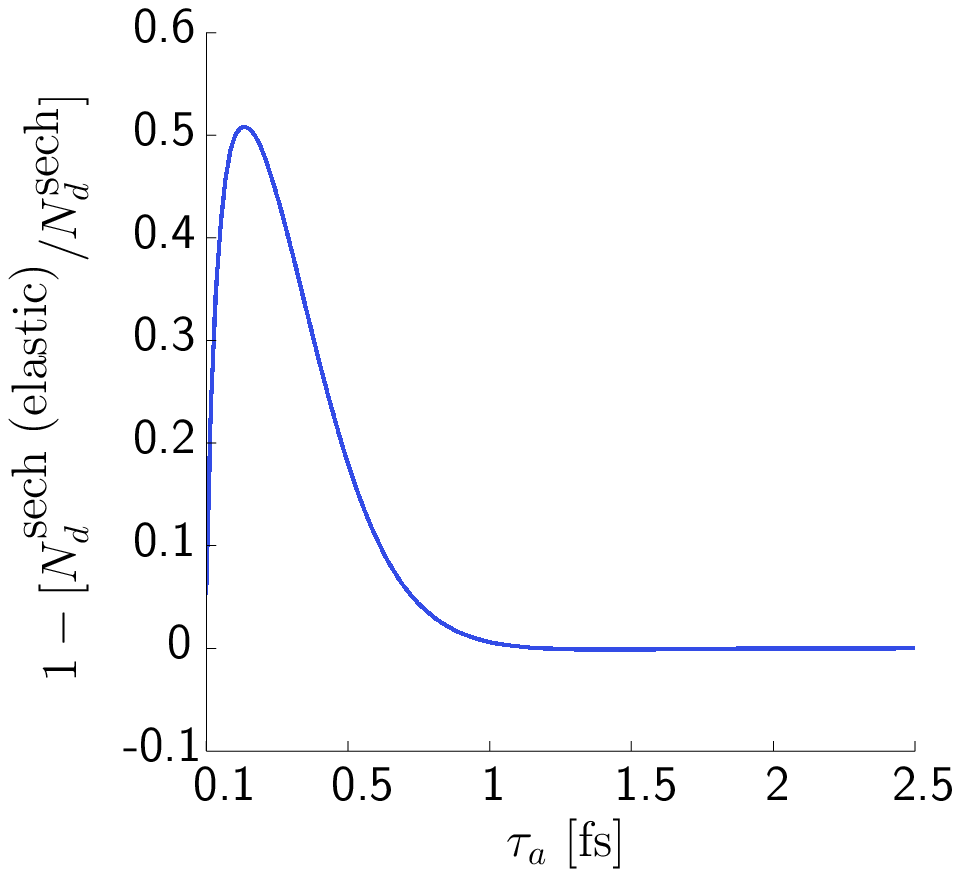}
\caption{On the left, for $N_{d}^{\trm{sech}}(\tau_{a})$, with
$\tau_{b}=1.4~\trm{fs}$, at $x_{d} = 0.1\, r_{d}$,  $z_{d}=0$, $y_{d} =
1~\trm{m}$ and $\lambda_{a} = \lambda_{b} = 0.8~\mu\trm{m}$, the dominant term
$E_{a}^{2}E_{b}$ has been plotted and the coefficient of the integral ignored.
On the right is plotted the relative difference between the ``elastic'' and full
behaviour of $N_{d}^{\trm{sech}}(\tau_{a})$.} \label{fig:IE_sech1} 
\end{figure}

The first plot in \figref{fig:IE_sech1} depicts the dependence of
$N^{\trm{(sech)}}_{d}$ on $\tau_{a}$ and we notice that for
$\omega_{a}\tau_{a}\lesssim2$ ($\tau_{a}\lesssim 0.8~\trm{fs}$), there is indeed
a deviation in the behaviour of $N^{\trm{(sech)}}_{d}$. We can take data from a
more uniform region $\tau_{a}>1~\trm{fs}$ and acquire a best-fit polynomial with
the boundary condition $N^{\trm{(sech)}}_{d}(\tau_{a}=0)=0$. It turns out that a
cubic polynomial fits the calculated points well (similar to the Gaussian beam
case where $N_{d}(\tau) \propto \tau^{3}$). When the fit parameters were
calculated for $1~\trm{fs}<\tau_{a}<2~\trm{fs}$, the goodness-of-fit was tested
with a Pearson's chi-squared test over $1~\trm{fs}<\tau_{a}<3~\trm{fs}$ and
found to support the hypothesis of agreement with a probability of over $0.995$.
When the relative difference of this ``elastic'' curve from the total was
calculated, the second plot in \figref{fig:IE_sech1} was generated. This clearly
demonstrates the new behaviour occurring for short pulse durations or
equivalently large bandwidths and so further supports our initial ansatz that
just one beam split into two counter-propagating sub-cycle pulses is sufficient
for accessing the process of vacuum inelastic photon-photon scattering. A suggestion for further work would be to investigate the role of the carrier-envelope phase as well as a chirped frequency.







\section{Summary}
In calculating numbers of photons scattered in the collision of two laser beams,
we had three aims: i) to consider a more realistic set-up of the colliding beams
(including a temporal pulse shape, collision angle, lag and lateral separation),
which would produce more accurate qualitative and quantitative predictions for
experiment, ii) to investigate the possibility of using a single laser, split
into two beams to measure elastic photon-photon scattering and iii) to evaluate
the ansatz that just two lasers, with sufficiently short pulse durations, can be
used to measure the process of inelastic photon-photon scattering. The first of
these aims has been met in \figref{fig:elastic_res} where the dependency on
various collision parameters was calculated and found consistent with physical
reasoning. This led to the second aim, where the inclusion of a pulse form and
collision angle led to two orders of magnitude difference over previous
elastic photon scattering estimates \cite{kingthesis} (the single-slit limit of \cite{king10a}). In this more
complete description, it was shown that when a $10~\trm{PW}$,
$\lambda=0.91~\trm{nm}$ beam is separated into two $30~\trm{fs}$ Gaussian
pulses, incident at an angle $0.1$, one could expect approximately $0.7$, $4$ or
$13$ photons, corresponding to the fundamental, second and fourth harmonic of
the probe respectively (with an assumed loss of $40\,\%$ per frequency
doubling), to be diffracted into detectable regions. As argued in
\cite{king10a}, this could be sufficient for measuring elastic photon-photon
scattering, here shown using a single $10~\trm{PW}$ source. The final aim was
partially met, first by considering Gaussian pulses, where it was shown that for
$\omega\tau_{a}\lesssim 4$ for the more intense beam $a$, the inelastic
scattering process increased and became as large as around $20$\% that of the
elastic count for $\omega\tau_{a}\approx 2$. However, for these results to be
consistent, $\omega\tau_{a}\gg 1$, so the head-on collision of two $\sech$
pulses was analysed, for which no such bound applies, where it was shown that
again, in this different field background, for $\omega\tau_{a}\approx 2$,
inelastic scattering became important --  as large as around $50$\% that of the
elastic one, lending supporting to our original ansatz.
\section{Acknowledgements}
B. K. would like to thank A. Di Piazza for his critical comments and careful reading of the manuscript.
\appendix
\renewcommand{\theequation}{\thesection.\arabic{equation}}
\section{Integration formulae}
\subsection{Gaussian diffracted field formulae}
Sum coefficients:
\begin{center}
\begin{table}[!h]
\begin{tabular}{r|rrrrrrrrrrrr}
$j$           & $\phantom{-}1$  & $2$ &  $3$  & $4$ & $5$ & $6$ & $\phantom{-}7$
& $8$ & $\phantom{-}9$ & $10$ & $11$ & $12$ \\
\hline
$A_{j}$       & $1$  & $1$ & $-2$  & $-1$ & $-1$ & $2$  & $1$ & $1$ & $-2$  &
$-1$ & $-1$ & $2$ \\
$\beta_{j}$   & $2$   &$-2$  & $0$  & $2$   & $-2$  &$0$  & $1$ & $1$ & $1$ &
$-1$ & $-1$ & $-1$\\
$\gamma_{j}$  & $1$  & $1$ &  $1$ & $-1$  & $-1$ & $-1$ & $2$  & $-2$  &$0$  &
$2$  & $-2$  & $0$ \\
\end{tabular}
\caption{Sum coefficients that occur in the expressions for $\mbf{E}_{d}$ and
$\widetilde{\mbf{E}}_{d}$, \eqnrefs{eqn:Ed}{eqn:Edtilde}}\label{tab:coeffs}
\end{table}
\end{center}
Diffracted field polarisation vectors:
\bea
\mbf{v}_{1} &=&
4\Big(\uveceps_{a}'\cdot\uveceps_{b}(1+\cos\theta)-\uvec{y}'\cdot\uveceps_{b}
~\uvec{y}\cdot\uveceps_{a}'\Big)\Big[\uveceps_{a}'\Big(1-(\stackrel{\leftarrow}{
\cdot}\uvec{r}~\uvec{r})\Big) + (\uvec{y}'\wedge\uveceps_{a}')\wedge\uvec{r}
\Big]\nonumber\\
&& +7\Big(\pmb{\eps}_{a}' \cdot \hat{\mbf{y}}\wedge \hat{\pmb{\eps}}_{b} -
\pmb{\eps}_{b} \cdot \hat{\mbf{y}}'\wedge \hat{\pmb{\eps}}_{a}'\Big)
\Big[-(\uvec{y}'\wedge\uveceps_{a}')
\Big(1-(\stackrel{\leftarrow}{\cdot}\uvec{r}~\uvec{r})\Big) +
\uveceps_{a}'\wedge\uvec{r}\Big]\label{eqn:v1}\\
\mbf{v}_{2} &=&
4\Big(\uveceps_{a}'\cdot\uveceps_{b}(1+\cos\theta)-\uvec{y}'\cdot\uveceps_{b}
~\uvec{y}\cdot\uveceps_{a}'\Big)\Big[\uveceps_{b}\Big(1-(\stackrel{\leftarrow}{
\cdot}\uvec{r}~\uvec{r})\Big) - (\uvec{y}'\wedge\uveceps_{b})\wedge\uvec{r}
\Big]\nonumber\\
&& +7\Big(\pmb{\eps}_{a}' \cdot \hat{\mbf{y}}\wedge \hat{\pmb{\eps}}_{b} -
\pmb{\eps}_{b} \cdot \hat{\mbf{y}}'\wedge \hat{\pmb{\eps}}_{a}'\Big)
\Big[(\uvec{y}\wedge\uveceps_{b})
\Big(1-(\stackrel{\leftarrow}{\cdot}\uvec{r}~\uvec{r})\Big) +
\uveceps_{b}\wedge\uvec{r}\Big]\label{eqn:v2}.
\eea
Integration terms:
\bea
I_{t,j} &=& \int d^{3}x\,
\frac{1}{[1+(y'/y_{r,a})^{2}]^{B_{j}/2}}\frac{1}{[1+(y/y_{r,b})^{2}]^{\Gamma_{j}
/2}}\nonumber \\
&&
\mbox{e}^{-x^{2}\big[\frac{B_{j}}{w_{a}^{2}(y')} +
\frac{\Gamma_{j}}{w_{p}^{2}(y)} + \frac{1}{\tau_{j}^{2}} +    			
i(\frac{\omega_{j}}{2r_{d}}(1-\frac{x_{d}^{2}}{r_{d}^{2}}) 			
			-\frac{\beta_{j}\omega_{a}y'}{2(y'^{2}+y^{2}_{r,a})} + 						
\frac{\gamma_{j}\omega_{b}y}{2(y^{2}+y_{r,b}^{2})}) \big] }\nonumber\\
&&
          \mbox{e}^{x \big[\frac{2x_{d}}{\tau_{j}^{2}} +
2\frac{x_{0}\Gamma_{j}}{w_{b}^{2}}
-2\frac{x_{d}}{r_{d}}(\frac{t}{\tau_{j}^{2}}+\frac{B_{j}(y'-\Delta
t)}{\tau_{a}^{2}} - \frac{\Gamma_{j}y}{\tau_{b}^{2}}) +
i(\omega_{j}(\frac{x_{d}}{r_{d}} + \frac{x_{d}(yy_{d}+zz_{d})}{r_{d}^{3}}) +
i\frac{\gamma_{j}\omega_{b}x_{0}y}{y^{2}+y_{r,b}^{2}}) \big]} \nonumber \\
&&         
\mbox{e}^{i\frac{\gamma_{j}\omega_{b}x_{0}^{2}y}{2(y^{2}+y_{r,b}^{2})} 
          + i\frac{\beta_{j}\omega_{a}y'z'^{2}}{2(y'^{2}+ y_{r,a}^{2})}  +
i\frac{\gamma_{j}\omega_{b}y(z-z_{0})^{2}}{2(y^{2}+y_{r,b}^{2})}
          + i(\omega_{a}\beta_{j}y'-\gamma_{j}\omega_{b}y -
\beta_{j}\tan^{-1}(y/y_{r,a}) + \gamma_{j} \tan^{-1} (y/y_{r,b})) } \nonumber\\
&&
          \mbox{e}^{- (\frac{B_{j}}{\tau_{a}^{2}}(y'-\Delta t)^{2} +
\frac{\Gamma_{j}y^{2}}{\tau_{b}^{2}})
 -\frac{1}{\tau_{j}^{2}}(t^{2}+y^{2}+z^{2}+r_{d}^{2}-2(yy_{d}+zz_{d}))
-i\omega_{j}(r_{d} -\frac{yy_{d}+zz_{d}}{r_{d}} + \frac{y^{2}+z^{2}}{2r_{d}} -
\frac{(yy_{d}+zz_{d})^{2}}{2r_{d}^{3}})
+ i\omega_{j}t }\nonumber \\ 
&& \mbox{e}^{-2t(\frac{B_{j}(y'-\Delta
t)}{\tau_{a}^{2}}-\frac{\Gamma_{j}y}{\tau_{b}^{2}})
+
2(r_{d}-\frac{yy_{d}+zz_{d}}{r_{d}})(\frac{t}{\tau_{j}^{2}}+\frac{B_{j}
(y'-\Delta t)}{\tau_{a}^{2}}-\frac{\Gamma_{j}y}{\tau_{b}^{2}}) -
\frac{B_{j}z'^{2}}{w_{a}^{2}}-\frac{\Gamma_{j}(z-z_{0})^{2}}{w_{b}^{2}} -
\frac{\Gamma_{j}x_{0}^{2}}{w_{b}^{2}} }.
\label{eqn:It}
\eea
\bea
I_{\omega,j} &=& \int d^{3} x\,\frac{1}{[1+(y'/y_{r,a})^{2}]^{B_{j}/2}}
\frac{1}{[1+(y/y_{r,b})^{2}]^{\Gamma_{j}/2}}\nonumber\\
&& \mbox{e}^{-\frac{B_{j}}{w_{a}^{2}}(x^{2}+z'^{2}) -
\frac{\Gamma_{j}}{w_{b}^{2}}[(x-x_{0})^{2}+(z-z_{0})^{2}] +
i(\beta_{j}\omega_{a}y' - \gamma_{j}\omega_{b}y) -
i\beta_{j}\tan^{-1}(y'/y_{r,a}) + i\gamma_{j}\tan^{-1}(y/y_{r,b})}\nonumber\\
&&
\mbox{e}^{\frac{i\beta_{j}\omega_{a}y'}{2}\frac{x^{2}+z'^{2}}{y'^{2}+y_{r,a}^{2}
} -
\frac{i\gamma_{j}\omega_{b}y}{2}\frac{(x-x_{0})^{2}+(z-z_{0})^{2}}{y^{2}+y_{r,b}
^{2}} - (\frac{B_{j}(y'-\Delta t)^{2}}{\tau_{a}^{2}} +
\frac{\Gamma_{j}y^{2}}{\tau_{b}^{2}}) - \frac{\omega^{2}\tau_{j}^{2}}{2} +
\tau_{j}^{2}(\frac{B_{j}(y'-\Delta
t)}{\tau_{a}^{2}}-\frac{\Gamma_{j}y}{\tau_{b}^{2}})^{2}+\frac{\omega\omega_{j}
\tau_{j}^{2}}{2} }\nonumber\\
&&\mbox{e}^{-i\omega(r_{d}-\frac{xx_{d}+yy_{d}+zz_{d}}{r_{d}} +
\frac{x^{2}+y^{2}+z^{2}}{2r_{d}} -
\frac{(xx_{d}+yy_{d}+zz_{d})^{2}}{2r_{d}^{3}})
+i(\omega-\omega_{j})\tau_{j}^{2}(\frac{B_{j}(y'-\Delta
t)}{\tau_{a}^{2}}-\frac{\Gamma_{j}y}{\tau_{b}^{2}})}.
\label{eqn:Iomega}
\eea

\subsection{Analytical approximation to $dN_{d}$}
\bea
dN_{d} &=& \sum_{p,q=1}^{12} dN_{d}^{pq} \\
dN_{d}^{pq} &=&  \frac{1}{\pi^{3/2}}
\Bigg[\frac{\alpha}{180}\frac{w_{a,0}^{2}}{r_{d}}\Bigg]^{2}
\frac{E_{a,0}^{B_{p}+B_{q}}E_{b,0}^{\Gamma_{p}+\Gamma_{q}}}{E_{cr}^{4}}\frac{
\mbf{v}_{l(p)}\cdot\mbf{v}_{l(q)}}{16^{2}}A_{p}A_{q}\tau_{p}^{2}\tau^{2}_{q} \\
&& \frac{1}{[B_{p} + \Gamma_{p}(\frac{w_{a,0}}{w_{b,0}})^{2}] [B_{q} +
\Gamma_{q}(\frac{w_{a,0}}{w_{b,0}})^{2}] }
\frac{1}{\sqrt{(1-\tau_{pp}^{4})(1-\tau_{qq}^{4})}}
 \frac{b(6a+b^{2})}{a^{7/2}}\trm{Erf}\Big(\frac{b}{2\sqrt{a}}\Big)\mbox{e}^{\frac{b^{2}}
{4a}+c}\nonumber\\
a & = & \frac{\rho_{d}^{2}w_{a,0}^{2}}{2sr_{d}^{2}} +
\frac{\tau_{p}^{2}}{4}\Bigg[ 1 + \frac{(y_{d}/r_{d} +
\tau_{qq}^{2})^{2}}{(1-\tau_{pp}^{4})} \Bigg] + (p\leftrightarrow q)\\
b & = & \frac{\tau_{p}^{2}}{2}\Bigg[\omega_{p} - \frac{(\tilde{\omega}_{p} - \omega_{p}\tau_{qq}^{2})(y_{d}/r_{d} +
\tau_{qq}^{2}) }{1-\tau_{pp}^{4}}\Bigg] + (p\leftrightarrow q)\\
c & = & -\frac{\tau^{2}_{p}}{4}\Bigg[\omega_{p}^{2} + \frac{(\tilde{\omega}_{p} - \omega_{p}\tau_{qq}^{2})^{2}}{1-\tau_{pp}^{4}} \Bigg] +
(p\leftrightarrow q)
\label{eqn:dN_analytic}
\eea

\subsection{Sech diffracted field formulae}
Integration terms:
\bea
I_{\omega,j}^{\trm{sech}} & = & \int_{0}^{\rho_{0}}\!\! d\rho\,\rho
\int^{2\pi}_{0}\!\! d\vphi \int^{\infty}_{-\infty}\!\! d\phi_{-}
\int^{\infty}_{-\infty}\!\! d\phi_{+}  ~\sech^{3-j}(\phi_{-}/\tau_{a})
\sech^{j}(\phi_{+}/\tau_{b}) \nonumber \\
&& \cos^{3-j}(\phi_{-}/\tau_{a})
\cos^{j}(\phi_{+}/\tau_{b})\mbox{e}^{\frac{i\omega}{2r_{d}}\rho^{2}(-1 +
\frac{x_{d}^{2}}{r_{d}^{2}}\cos^{2}\!\vphi) +
i\omega\rho\frac{x_{d}y_{d}}{2r_{d}^{3}}(\phi_{+}-\phi_{-})\cos\vphi}\nonumber\\
&& \mbox{e}^{i\omega\rho\frac{x_{d}}{r_{d}}\cos\vphi +\frac{i\omega}{2}
(\phi_{-}(1-\frac{y_{d}}{r_{d}}) +
\phi_{+}(1+\frac{y_{d}}{r_{d}}))-\frac{i\omega}{8r_{d}}(\phi_{-}-\phi_{+})^{2}
(1-\frac{y_{d}^{2}}{r_{d}^{2}})} \label{eqn:Isech}
\eea

\providecommand{\noopsort}[1]{}


\begin{thebibliography}{25}
\expandafter\ifx\csname natexlab\endcsname\relax\def\natexlab#1{#1}\fi
\expandafter\ifx\csname bibnamefont\endcsname\relax
  \def\bibnamefont#1{#1}\fi
\expandafter\ifx\csname bibfnamefont\endcsname\relax
  \def\bibfnamefont#1{#1}\fi
\expandafter\ifx\csname citenamefont\endcsname\relax
  \def\citenamefont#1{#1}\fi
\expandafter\ifx\csname url\endcsname\relax
  \def\url#1{\texttt{#1}}\fi
\expandafter\ifx\csname urlprefix\endcsname\relax\def\urlprefix{URL }\fi
\providecommand{\bibinfo}[2]{#2}
\providecommand{\eprint}[2][]{\url{#2}}

\bibitem[{\citenamefont{Odom \emph{et al.}}(2006)}]{gabrielse06}
\bibinfo{author}{\bibfnamefont{B.}~\bibnamefont{Odom}} \bibnamefont{\emph{et
  al.}}, \bibinfo{journal}{Phys. Rev. Lett.} \textbf{\bibinfo{volume}{97}},
  \bibinfo{eid}{030801} (\bibinfo{year}{2006}).

\bibitem[{\citenamefont{Sturm \emph{et al.}}(2011)}]{sturm11}
\bibinfo{author}{\bibfnamefont{S.}~\bibnamefont{Sturm}} \bibnamefont{\emph{et
  al.}}, \bibinfo{journal}{Phys. Rev. Lett.} \textbf{\bibinfo{volume}{107}},
  \bibinfo{pages}{023002} (\bibinfo{year}{2011}).

\bibitem[{\citenamefont{Sauter}(1931)}]{sauter31}
\bibinfo{author}{\bibfnamefont{F.}~\bibnamefont{Sauter}}, \bibinfo{journal}{Z.
  Phys.} \textbf{\bibinfo{volume}{69}}, \bibinfo{pages}{742}
  (\bibinfo{year}{1931}).

\bibitem[{\citenamefont{Heisenberg and Euler}(1936)}]{heisenberg_euler36}
\bibinfo{author}{\bibfnamefont{W.}~\bibnamefont{Heisenberg}} \bibnamefont{and}
  \bibinfo{author}{\bibfnamefont{H.}~\bibnamefont{Euler}}, \bibinfo{journal}{Z.
  Phys.} \textbf{\bibinfo{volume}{98}}, \bibinfo{pages}{714}
  (\bibinfo{year}{1936}).

\bibitem[{\citenamefont{Ferrando \emph{et al.}}(2007)}]{tommasini07}
\bibinfo{author}{\bibfnamefont{A.}~\bibnamefont{Ferrando}}
  \bibnamefont{\emph{et al.}}, \bibinfo{journal}{Phys. Rev. Lett.}
  \textbf{\bibinfo{volume}{99}}, \bibinfo{pages}{150404}
  (\bibinfo{year}{2007}).

\bibitem[{\citenamefont{Mendonca \emph{et al.}}(2006)}]{mendonca06}
\bibinfo{author}{\bibfnamefont{J.~T.} \bibnamefont{Mendonca}}
  \bibnamefont{\emph{et al.}}, \bibinfo{journal}{Phys. Lett. A}
  \textbf{\bibinfo{volume}{359}}, \bibinfo{pages}{700} (\bibinfo{year}{2006}).

\bibitem[{\citenamefont{King \emph{et
  al.}}(2010{\natexlab{a}})\citenamefont{King, Piazza, and Keitel}}]{king10b}
\bibinfo{author}{\bibfnamefont{B.}~\bibnamefont{King}},
  \bibinfo{author}{\bibfnamefont{A.~D.} \bibnamefont{Piazza}},
  \bibnamefont{and} \bibinfo{author}{\bibfnamefont{C.~H.}
  \bibnamefont{Keitel}}, \bibinfo{journal}{Phys. Rev. A}
  \textbf{\bibinfo{volume}{82}}, \bibinfo{pages}{032114}
  (\bibinfo{year}{2010}{\natexlab{a}}).

\bibitem[{\citenamefont{{Di Piazza} \emph{et al.}}(2006)\citenamefont{{Di
  Piazza}, Hatsagortsyan, and Keitel}}]{dipiazza_PRL_06}
\bibinfo{author}{\bibfnamefont{A.}~\bibnamefont{{Di Piazza}}},
  \bibinfo{author}{\bibfnamefont{K.~Z.} \bibnamefont{Hatsagortsyan}},
  \bibnamefont{and} \bibinfo{author}{\bibfnamefont{C.~H.}
  \bibnamefont{Keitel}}, \bibinfo{journal}{Phys. Rev. Lett.}
  \textbf{\bibinfo{volume}{97}}, \bibinfo{pages}{083603}
  (\bibinfo{year}{2006}).

\bibitem[{\citenamefont{Heinzl \emph{et al.}}(2006)}]{heinzl_birefringence06}
\bibinfo{author}{\bibfnamefont{T.}~\bibnamefont{Heinzl}} \bibnamefont{\emph{et
  al.}}, \bibinfo{journal}{Opt. Commun.} \textbf{\bibinfo{volume}{267}},
  \bibinfo{pages}{318} (\bibinfo{year}{2006}).

\bibitem[{\citenamefont{Homma \emph{et al.}}(2011)\citenamefont{Homma, Habs,
  and Tajima}}]{tajima11a}
\bibinfo{author}{\bibfnamefont{K.}~\bibnamefont{Homma}},
  \bibinfo{author}{\bibfnamefont{D.}~\bibnamefont{Habs}}, \bibnamefont{and}
  \bibinfo{author}{\bibfnamefont{T.}~\bibnamefont{Tajima}},
  \bibinfo{journal}{Appl. Phys. B-Lasers O.} \textbf{\bibinfo{volume}{104}},
  \bibinfo{pages}{769} (\bibinfo{year}{2011}).

\bibitem[{\citenamefont{King \emph{et
  al.}}(2010{\natexlab{b}})\citenamefont{King, Piazza, and Keitel}}]{king10a}
\bibinfo{author}{\bibfnamefont{B.}~\bibnamefont{King}},
  \bibinfo{author}{\bibfnamefont{A.~D.} \bibnamefont{Piazza}},
  \bibnamefont{and} \bibinfo{author}{\bibfnamefont{C.~H.}
  \bibnamefont{Keitel}}, \bibinfo{journal}{Nature Photon.}
  \textbf{\bibinfo{volume}{4}}, \bibinfo{pages}{92}
  (\bibinfo{year}{2010}{\natexlab{b}}).

\bibitem[{\citenamefont{Kryuchkyan and Hatsagortsyan}(2011)}]{hatsagortsyan11}
\bibinfo{author}{\bibfnamefont{G.~Y.} \bibnamefont{Kryuchkyan}}
  \bibnamefont{and} \bibinfo{author}{\bibfnamefont{K.~Z.}
  \bibnamefont{Hatsagortsyan}}, \bibinfo{journal}{Phys. Rev. Lett.}
  \textbf{\bibinfo{volume}{107}}, \bibinfo{pages}{053604}
  (\bibinfo{year}{2011}).

\bibitem[{\citenamefont{{Di Piazza} \emph{et al.}}(2005)\citenamefont{{Di
  Piazza}, Hatsagortsyan, and Keitel}}]{dipiazza_harmonic05}
\bibinfo{author}{\bibfnamefont{A.}~\bibnamefont{{Di Piazza}}},
  \bibinfo{author}{\bibfnamefont{K.~Z.} \bibnamefont{Hatsagortsyan}},
  \bibnamefont{and} \bibinfo{author}{\bibfnamefont{C.~H.}
  \bibnamefont{Keitel}}, \bibinfo{journal}{Phys. Rev. D}
  \textbf{\bibinfo{volume}{72}}, \bibinfo{pages}{085005}
  (\bibinfo{year}{2005}).

\bibitem[{\citenamefont{Yanovsky \emph{et al.}}(2008)}]{yanovsky08}
\bibinfo{author}{\bibfnamefont{V.}~\bibnamefont{Yanovsky}}
  \bibnamefont{\emph{et al.}}, \bibinfo{journal}{Opt. Express}
  \textbf{\bibinfo{volume}{16}}, \bibinfo{pages}{2109} (\bibinfo{year}{2008}).

\bibitem[{\citenamefont{Vulcan}(2010)}]{vulcan_site}
\bibinfo{author}{\bibfnamefont{C.~F.} \bibnamefont{Vulcan}},
  \emph{\bibinfo{title}{Vulcan glass laser}},
  \bibinfo{howpublished}{http://www.clf.rl.ac.uk/Facilities/vulcan/index.htm}
  (\bibinfo{year}{2010}).

\bibitem[{\citenamefont{Tommasini and Michinel}(2010)}]{tommasini10}
\bibinfo{author}{\bibfnamefont{D.}~\bibnamefont{Tommasini}} \bibnamefont{and}
  \bibinfo{author}{\bibfnamefont{H.}~\bibnamefont{Michinel}},
  \bibinfo{journal}{Phys. Rev. A (R)} \textbf{\bibinfo{volume}{82}},
  \bibinfo{pages}{011803} (\bibinfo{year}{2010}).

\bibitem[{\citenamefont{Brodin \emph{et al.}}(2001)\citenamefont{Brodin,
  Marklund, and Stenflo}}]{marklund_PRL_01}
\bibinfo{author}{\bibfnamefont{G.}~\bibnamefont{Brodin}},
  \bibinfo{author}{\bibfnamefont{M.}~\bibnamefont{Marklund}}, \bibnamefont{and}
  \bibinfo{author}{\bibfnamefont{L.}~\bibnamefont{Stenflo}},
  \bibinfo{journal}{Phys. Rev. Lett.} \textbf{\bibinfo{volume}{87}},
  \bibinfo{pages}{171801} (\bibinfo{year}{2001}).

\bibitem[{\citenamefont{Lundstr{\"o}m \emph{et al.}}(2006)}]{lundstroem_PRL_06}
\bibinfo{author}{\bibfnamefont{E.}~\bibnamefont{Lundstr{\"o}m}}
  \bibnamefont{\emph{et al.}}, \bibinfo{journal}{Phys. Rev. Lett.}
  \textbf{\bibinfo{volume}{96}}, \bibinfo{pages}{083602}
  (\bibinfo{year}{2006}).

\bibitem[{\citenamefont{Fedotov and Narozhny}(2006)}]{fedotov_harmonics06}
\bibinfo{author}{\bibfnamefont{A.~M.} \bibnamefont{Fedotov}} \bibnamefont{and}
  \bibinfo{author}{\bibfnamefont{N.~B.} \bibnamefont{Narozhny}},
  \bibinfo{journal}{Phys. Lett. A} \textbf{\bibinfo{volume}{362}},
  \bibinfo{pages}{1} (\bibinfo{year}{2006}).

\bibitem[{\citenamefont{{Y. I. Salamin} \emph{et
  al.}}(2006)}]{salamin_review06}
\bibinfo{author}{\bibnamefont{{Y. I. Salamin}}} \bibnamefont{\emph{et al.}},
  \bibinfo{journal}{Phys. Rep.} \textbf{\bibinfo{volume}{427}},
  \bibinfo{pages}{41} (\bibinfo{year}{2006}).

\bibitem[{\citenamefont{{GNU Project - Free Software Foundation}}(2011)}]{gsl}
\bibinfo{author}{\bibnamefont{{GNU Project - Free Software Foundation}}},
  \emph{\bibinfo{title}{{GNU Scientific Library}}},
  \bibinfo{howpublished}{http://www.gnu.org/s/gsl/} (\bibinfo{year}{2011}).

\bibitem[{\citenamefont{Jackson}(1975)}]{jackson75}
\bibinfo{author}{\bibfnamefont{J.~D.} \bibnamefont{Jackson}},
  \emph{\bibinfo{title}{Classical Electrodynamics}} (\bibinfo{publisher}{John
  Wiley \& Sons, Inc.}, \bibinfo{address}{New York}, \bibinfo{year}{1975}).

\bibitem[{\citenamefont{King}(2010)}]{kingthesis}
\bibinfo{author}{\bibfnamefont{B.}~\bibnamefont{King}}, Ph.D. thesis,
  \bibinfo{school}{University of Heidelberg} (\bibinfo{year}{2010}).

\bibitem[{\citenamefont{Davis and Rabinowitz}(1967)}]{rabinowitz67}
\bibinfo{author}{\bibfnamefont{P.~J.} \bibnamefont{Davis}} \bibnamefont{and}
  \bibinfo{author}{\bibfnamefont{P.}~\bibnamefont{Rabinowitz}},
  \emph{\bibinfo{title}{Numerical integration}} (\bibinfo{publisher}{Blaisdell
  Publishing Company}, \bibinfo{address}{Waltham, Massachusetts},
  \bibinfo{year}{1967}).

\bibitem[{\citenamefont{Enge \emph{et al.}}(2011)}]{MPC}
\bibinfo{author}{\bibfnamefont{A.}~\bibnamefont{Enge}} \bibnamefont{\emph{et
  al.}}, \emph{\bibinfo{title}{{MPC}}},
  \bibinfo{howpublished}{http://www.multiprecision.org/}
  (\bibinfo{year}{2011}).

\end{thebibliography}
\end{document}